\documentclass[fleqn,usenatbib]{mnras}

\usepackage[T1]{fontenc}

\DeclareRobustCommand{\VAN}[3]{#2}
\let\VANthebibliography\thebibliography
\def\thebibliography{\DeclareRobustCommand{\VAN}[3]{##3}\VANthebibliography}

\usepackage{graphicx}	
\usepackage{amsmath}	
\usepackage{amssymb}	
\usepackage{pdflscape}

\usepackage{newtxtext,newtxmath}

\newcommand{\oiii}{[O{\sc iii}]\,\ensuremath{88.36\micron}}
\newcommand{\otwop}{\ensuremath{\mathrm{O}^{2+}}}

\title[Far-IR spectroscopy of $\gamma$ Vel]{Oxygen abundance of $\gamma$ Vel from [O\,{\sc iii}] 88$\mu$m  {\it Herschel}/PACS spectroscopy}

\author[Crowther et al.]{
Paul A. Crowther$^{1}$\thanks{paul.crowther@sheffield.ac.uk}, 
M. J. Barlow$^{2}$, 
P. Royer$^{3}$,
D. J. Hillier$^{4}$,
J. M. Bestenlehner$^{1}$,
P. W. Morris$^{5}$,
R. Wesson$^{6}$\\
$^{1}$ Department of Physics and Astronomy, University of Sheffield, Sheffield, S3 7RH, UK\\
$^{2}$  Department of Physics and Astronomy, University College London, 
Gower Street, London, WC1E 6BT, United Kingdom\\
$^{3}$ Institute of Astronomy, Celestijnenlaan 200d - box 2401, 3001 
Leuven, Belgium\\ 
$^{4}$ Department of Physics and Astronomy \& Pittsburgh Particle Particle Physics, Astrophysics and Cosmology Center (PITT PACC),\\
University of Pittsburgh, 100 Allen Hall, 391 O'Hara St, Pittsburgh PA 15260, USA\\
$^{5}$ IPAC M/C 100-22, California Institute 
of Technology, 770 S.Wilson Ave, Pasadena, CA 91125, USA\\
$^{6}$ School of Physics and Astronomy, Cardiff University, Queen's Buildings North Building, 5 The Parade, Newport Road, Cardiff, CF24 3AA, UK 
}

\date{Accepted 2024 XX. Received 2024 XX; in original form 2023 XX}

\pubyear{2024}

\begin{document}
\label{firstpage}
\pagerange{\pageref{firstpage}--\pageref{lastpage}}
\maketitle

\begin{abstract}
We present {\it Herschel} PACS spectroscopy of the [O\,{\sc iii}] 88.4$\mu$m fine-structure line in the nearby WC8+O binary system $\gamma$ Vel to determine its oxygen abundance. The critical density of this
line corresponds to several 10$^{5} R_{\ast}$ such that it is spatially extended in PACS observations at the 336~pc distance to $\gamma$ Vel. Two approaches
are used, the first involving a detailed stellar atmosphere analysis of $\gamma$ Vel using {\sc cmfgen}, extending to $N_{e} \sim 10^{0}$ cm$^{-3}$ in order to fully sample the line formation region of [O\,{\sc iii}] 88.4$\mu$m. The second approach involves the analytical model introduced by Barlow et al. and revised by Dessart et al., additionally exploiting {\it ISO} LWS spectroscopy of [O\,{\sc iii}] 51.8$\mu$m. We obtain higher luminosities for the WR and O components of $\gamma$ Vel with respect to De Marco et al., $\log L/L_{\odot}$ = 5.31 and 5.56, respectively, primarily due to the revised  (higher) interferometric distance.
We obtain an oxygen mass fraction of $X_{\rm O}$ = 1.0$\pm$0.3\% for an outer wind volume filling factor of $f$ = 0.5$\pm$0.25, favouring either standard or slightly reduced Kunz et al. rates for the $^{12}\mathrm{C}({\alpha}, {\gamma})^{16}\mathrm{O}$ reaction from comparison with BPASS binary population synthesis models. We also revisit neon and sulphur abundances in the outer wind of $\gamma$ Vel from {\it ISO} SWS spectroscopy of [S\,{\sc iv}] 10.5$\mu$m, [Ne\,{\sc ii}] 12.8$\mu$m and [Ne\,{\sc iii}] 15.5$\mu$m. The inferred neon abundance $X_{\rm Ne}$ = 2.0$_{-0.6}^{+0.4}$\% is in excellent agreement with BPASS predictions, while the sulphur abundance of $X_{\rm S}$ = 0.04$\pm$0.01\% agrees with the solar abundance, as expected for unprocessed elements.
\end{abstract}

\begin{keywords}
stars: Wolf-Rayet - stars: early-type - stars: massive - stars: abundances - infrared: stars
\end{keywords}



\section{Introduction}

The majority of oxygen and other $\alpha$-elements in the universe are created
in massive stars. Such stars not only achieve requisite temperatures to 
synthesise heavy elements, but during late evolutionary phases inject immense 
mechanical energy to mix these fresh ingredients into the
interstellar medium (ISM). Wolf-Rayet (WR) stars are the evolved descendants of
O-stars which exhibit prodigious mass-loss rates of 
$\sim 10^{-5}M_\odot$ yr$^{-1}$ via line-driven winds, and are characterised by
strong emission line spectra of nitrogen (WN), carbon (WC) or oxygen
(WO), representing a broad evolutionary sequence \citep{crowther07}. Evidence
for chemical enrichment of the ISM by WR winds has been observed in the
circumstellar nebulae of individual WR stars \citep{esteban92,stock11}, and
invoked to explain an increased N/O ratio in galaxies displaying WR spectral
features \citep{brinchmann08,lopez-sanchez10}.

WC (and  WO) stars represent the final evolutionary stages of 
$M_{ini}\,{\gtrsim}\,20M_\odot$ stars \citep{meynet05}. Following a lifetime 
of extreme mass-loss, these `naked' helium stars offer a unique window into the 
chemical evolution of massive star cores. Accurate measurement of abundances 
in such stars stars can therefore provide important constraints on nuclear 
reaction rates and the efficiency of various internal mixing processes.

The precise amount of oxygen produced in massive stars has long been a contentious 
issue \citep{fowler84}. The ratio $^{12}\mathrm{C}/^{16}\mathrm{O}$ is sensitive to  
the core density and temperature and duration of Helium burning, but it depends 
directly on the competition in rates of the $3\alpha\,{\rightarrow}\,^{12}\mathrm{C}$  and  
$^{12}\mathrm{C}({\alpha}, {\gamma})^{16}\mathrm{O}$ processes.
The rate of $3\alpha\,{\rightarrow}\,^{12}\mathrm{C}$ is reasonably secure
\citep{fynbo05}, but a significant level of uncertainty remains regarding the
rate of $^{12}\mathrm{C}({\alpha}, {\gamma})^{16}\mathrm{O}$  \citep{Buchmann06}.
\citet{Aadland22} have recently discussed implications of varying rates for contemporary evolutionary models of massive stars. 

The abundances of helium and carbon in WC atmospheres are well known. Model
atmosphere codes such as {\sc cmfgen} \citep{hillier98} and PoWR \citep{grafener02}
have been developed to solve the transfer equation under non-LTE conditions in
an expanding, spherically symmetric, clumpy atmosphere. These have been applied 
to reproduce the plethora of carbon recombination lines in WC
near-UV/optical/near-IR spectra, typically revealing
$\mathrm{C/He}\,{=}\,0.1\mbox{--}0.4$ by number ($X_C\,{=}$ 20-50\%  by mass)
\citep{crowther02,sander12,Aadland22}. 

Despite the prevalence of oxygen lines in WC and particularly WO spectra, 
the measurement of abundances by spectroscopic modelling has been challenging, 
since C lines dominate most regions of the UV--optical spectrum, and many O
features are highly sensitive to ionisation \citep{crowther02}. 
The O lines most appropriate and commonly used 
to determine abundances in WC spectra are found in the near-UV
(3000--3500\AA), and are therefore severely affected by dust extinction
in the Galactic plane. Using space-based UV observations of WC and WO stars in
the LMC, \citet{Aadland22} measure oxygen abundances of
$\mathrm{O/He}\,{=}\,0.01\mbox{--}0.05$ by number
($X_O\,{=}$ 2--9\% by mass).
Another problem has been the complexity of model
atoms required to accurately compute radiative transfer solutions in the winds
of WC and WO stars, particularly for late-type WC stars abundant in 
C$^+$~\&~C$^{2+}$. \citet{williams15} presented O abundance
determinations in WC9 stars, based on {\sc cmfgen} model atmosphere analyses of optical spectra, 
obtaining $\mathrm{O/He}\,{=}\,\mbox{0.005--0.017} (X_O\,{=}$ 2-4\% by mass).



A more straightforward indicator of elemental abundance is provided by forbidden
fine-structure lines which originate at low electron densities, corresponding to very large radii
${>}\,10^3R_\ast$, where $R_\ast$ is the radius with Rosseland optical depth 
${\sim}10$. The flux in these lines is directly proportional to the
fractional abundance ($\gamma_i$) of the emitting ionic species 
\citep{barlow88}. However, this method  requires knowledge of the mass-loss properties and ionization structure. This 
approach has been 
adapted for clumped winds by \citet{dessart00}, who use mid-IR {\it ISO} SWS 
spectroscopy of neon and sulphur lines to calculate ionic abundances of these
elements in WC stars \citep[see also][]{morris00}. The mid-IR {\it Spitzer} IRS spectrograph has also
been used to study ionic abundances in WN and WC stars \citep{Morris04, crowther06}.

Fine splitting in the ground state of $\mathrm{O}^{2+}$ generates the
[OIII]$\,88.36\mu$m ($^3\mathrm{P}_1-^3\mathrm{P}_0$) forbidden fine-structure
line. With a critical density\footnote{$N_{\rm cric}$ is the density at which an energy level is depopulated equally by collisions and spontaneous emission \citep{2006agna.book.....O}.}  of $N_{\rm crit} {\sim}500\,\mathrm{cm}^{-3}$ \citep{Rubin89}, this line is
expected to originate from radii of a few ${\times}\,10^5 R_\ast$ in WC stars. 
\textit{Herschel}'s
Photoconductor Array Camera and Spectrometer (PACS) permits observation of Galactic
WC stars at  [OIII]$\,88.36\mu$m, from which we aim to provide an independent oxygen abundance.  

This paper is focused on [O\,{\sc iii}] 88.4$\mu$m PACS observations of $\gamma$ Vel (WR11, HD~68273)\footnote{This system is commonly referred to as $\gamma^{2}$ Vel, but at V=1.83 mag, is significantly brighter than $\gamma^{1}$ Vel (HD~68243), another early-type binary with V=4.17 mag, so we favour $\gamma$ Vel or $\gamma$ Vel A for the WR+O system}. At an interferometric distance of 336$^{+8}_{-7}$~pc \citep{North07, vanLeeuwen07}, 
this is comfortably the closest Wolf-Rayet star \citep{Rate20}. $\gamma$ Vel is a WC8+O binary system with a 78.5 day \citep{niemela80}, moderately eccentric orbit \citep{North07}. It has been extensively studied across the electromagnetic spectrum including X-rays \citep{2001ApJ...558L.113S, Schild04}, UV  \citep{1970ApJ...159..543S, stlouis93}, optical \citep{Schmutz97}, near-infrared \citep{Aitken82}, mid-infrared \citep{barlow88, vanderHucht96} and mm/radio \citep{Seaquist76, Williams90, 1999ApJ...518..890C}.  \citet{demarco99} and \citet{demarco00} have determined dynamical masses for its components of 30$M_{\odot}$ (O) and 9$M_{\odot}$ (WC8), adopting the original {\it Hipparcos} distance of $\sim$258 pc \citep{1997ApJ...484L.153S, deZeeuw99}. Its age is 5.5$\pm$1 Myr \citep{eldridge09} and is located within a self-named stellar cluster within the Vela OB2 association \citep{Jeffries09, Jeffries17, Franciosini18}.

In order to determine the oxygen abundance from fine-structure lines, it is necessary to determine the
outer WC wind properties. It is firmly established that the inner winds ($\mbox{1--10}\,R_\ast$) 
of hot, luminous stars are clumped, likely as a consequence of intrinsic instability in the line driving mechanism
\citep{owocki84}. This explains many observational features unaccounted for by
homogeneous wind models, such as the strength of electron scattering wings in WR
spectra \citep{hillier91}. For WR stars, consistency with observations has most
commonly been found for inner wind clumping with volume filling factor
$f\,{\simeq}\,0.1$ \citep{morris00,kurosawa02}. However, the radial extent
to which clumping persists and whether the filling factor remains constant is
unknown, but there is both theoretical and observational evidence to suggest
that it varies with radius.

\citet{runacres02} performed 1D simulations of a
line-driven hot-star wind out to $100\,R_\ast$. A highly structured wind is
generated, consisting of strong reverse shocks and weaker forward shocks
confining high density regions, separated by high-speed rarefied material. They
find that collisions of these dense regions allow structure to persist up to the
largest radii considered, with the so-called clumping factor 
($1/f$) rising to a maximum of ${\sim}\,10$ around $20\,R_\ast$
and falling to ${\sim}\,5$ at and beyond $50\,R_\ast$. As an
extension to these simulations, \citet{runacres05} take advantage
of the diminishing radiative force beyond ${\sim}\,30\,R_\ast$ to analyse the
evolution wind structure out to $1300\,R_\ast$ in a purely hydrodynamical
sense. Once again, persistent structure is seen out to the largest radii,
characterised by a clumping factor ${\sim}\,4~(f\,\sim\,0.25$) between
$300\mbox{--}1300\,R_\ast$. \citet{Sundqvist18} extend simulations to 2D, revealing somewhat lower
clumping factors. 


Variations in wind clumping are also observationally based, by the discrepancy in 
mass-loss rates derived using diagnostics originating from different depths in 
a stellar wind. For example, \citet{puls06} reveal a factor of two difference 
between H$\alpha$ derived $\dot{M}$ ($\leq 2\,R_\ast$) and radio derived $\dot{M}$ 
(${\sim}\,100\,R_\ast$) in O supergiants under the assumption of constant 
clumping factor, implying a lower clumping factor in the radio emitting region. \citet{Najarro11}
utilised mid-IR spectroscopy of O stars to quantify the radial dependence of clumping,
obtaining $f$ = 0.17 for $\zeta$ Pup (O4\,If) at $\sim 100\,R_\ast$, while \citet{RubioDiez22} 
favour volume filling factors of $f$ = 0.25--0.5 for OB supergiants from  far-IR photometry.

 In Section \ref{obs} we provide details of the 
observational datasets used. Section~\ref{cmfgen} details a spectroscopic analysis of $\gamma$ Vel
including fits to mid- and far-IR fine structure lines of oxygen, neon and sulphur. We summarise the analytical method of \citet{barlow88} in Section~\ref{BRA88} and apply this to $\gamma$ Vel in order to 
calculate  oxygen, neon and sulphur abundances, informed by our spectroscopic analysis. Comparisons between our derived oxygen abundance
and evolutionary predictions are provided in Section~\ref{summary} and  brief conclusions are drawn.        

\begin{figure}
\centering
\includegraphics[width=0.3\textwidth,angle=-90,bb=65 96 552 779]{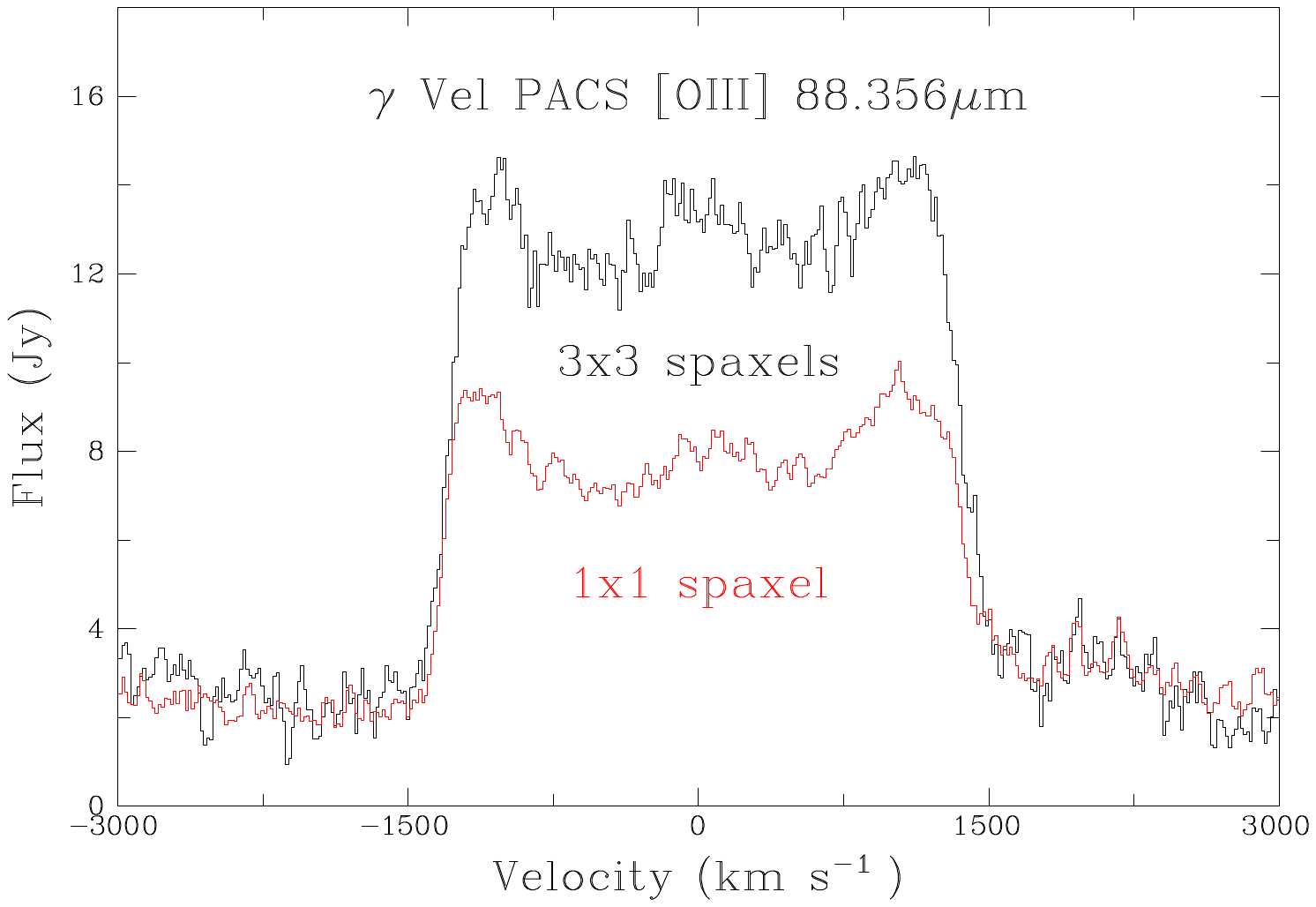}
\caption{PACS spectroscopy of [O\,{\sc iii}] 88.4$\mu$m fine structure line in $\gamma$ Vel, extracted from
3$\times$3 central spaxels (black) versus central spaxel, highlighting its spatial extent ${\gtrsim}\,1.5$ spaxels, 
i.e, ${\gtrsim}$\,4500 AU.} 
\label{oiii}
\end{figure}

\section{Observations}\label{obs}

\subsection{{\it Herschel} PACS spectroscopy}

$\gamma$ Vel was observed with the Photodetector Array Camera and Spectrometer \citep[PACS,][]{2010A&A...518L...2P} instrument on-board the
\textit{Herschel Space Telescope} on 7 Dec 2012 in spectroscopy mode (P.I. Crowther). 
PACS images a field of view of $47^{\prime\prime}\times47^{\prime\prime}$, resolved into a grid of
$5\times$5 spectroscopic pixels, or `spaxels', each spanning 9.4$^{\prime\prime}\times 9.4^{\prime\prime}$.
This field of view is rearranged via an image slicer onto two $16\times25$
Ge:Ga detector arrays, providing simultaneous  first order $\mbox{105--210}\micron$ ($R\sim$1000--2000) and
second order $\mbox{55--105}\micron$ ($R\sim$1500-3000) spectroscopy.

Scans totalling 2\,383\,s covering $\mbox{86.2--90.1\micron}$ with high spectral sampling density 
were carried out for $\gamma$ Vel. This wavelength range corresponds to 
${\pm}5000$\,kms$^{-1}$. The 88$\mu$m line was observed in second order ($R\sim$2500) in 
chopping/nodding mode with a small chopper throw, and an additional window was 
simultaneously observed at $176\mu$m in first order. The first (second) order datasets were 
reduced using {\it Herschel} Interactive Processing Environment (HIPE) 10.0.1496 (10.0.1297) and calibration set 45. We apply standard point source corrections (PSC) to the 
central spaxel in both instances, but prefer the integral over the central 9 spaxels for the second order dataset (excluding PSC correction) as discussed below.



For our adopted distance of 336$^{+8}_{-7}$~pc to $\gamma$ Vel, the $9.4^{\prime\prime}$ spaxel scale of PACS corresponds to
 $\sim$3000~AU. The critical density of 88$\mu$m  occurs at a radius of 
$2 \times 10^5R_\ast$ (${\sim}$\,2000 AU) for $R_\ast \sim 1.7 R_{\odot}$ (Section~\ref{cmfgen}). Therefore, the physical scale of
the emitting region, may be resolvable by PACS for $\gamma$ Vel. Indeed, in Figure~\ref{oiii} 
we show the PACS scan, revealing 88.4$\mu$m line emission extending beyond the central spaxel, confirming the physical 
extension of the emitting region as ${\gtrsim}\,1.5$ spaxels, i.e, 4500 AU. The total [O\,{\sc iii}] 88.4$\mu$m line
flux is $(3.4 \pm 0.5) \times 10^{-12}$ erg\,s$^{-1}$\,cm$^{-2}$, accounting for the 15\% absolute flux calibration uncertainties of PACS\footnote{https://www.cosmos.esa.int/documents/12133/996891/PACS+Explanatory+Supplement}. 
It also displays a measurable continuum flux, 
both adjacent to [O\,{\sc iii}] in second order (2.4$\pm$0.6 Jy) and at $176\mu$m in first order (1.5$\pm$0.3 Jy). As anticipated, 
no spectral features are detected in first order. \citet{Roche12} have previously spatially resolved the outer wind of $\gamma$ Vel from ground-based 
mid-IR observations of [S\,{\sc iv}] and [Ne\,{\sc ii}] fine structure lines.

Forbidden fine-structure lines  provide the most direct means of measuring terminal wind
velocities in WR stars, as they originate entirely from the asymptotically
flowing region of the stellar wind \citep{barlow88}. Since the line is
optically thin, it is expected to have a flat-topped and rectangular profile. [O\,{\sc iii}] 88.4$\mu$m exhibits structure within the flat-topped profile, although this is
reminiscent of [Ne\,{\sc iii}] 15.5$\mu$m observations of $\gamma$ Vel from {\it ISO}-SWS \citep{dessart00}. The full width of the emission line should therefore correspond to twice
$v_\infty$. However, in reality, the line profile will be modified by the
instrumental profile, making emission lines slightly non-rectangular. We adopt 
the Full Width at Zero Intensity (FWZI) as twice $v_\infty$.  We obtain $v_{\infty}$ = 1500$\pm$20 km\,s$^{-1}$
from [O\,{\sc iii}] 88.36$\mu$m, in good agreement with previous fine-structure \citep{barlow88}, near-IR \citep{Eenens94} and UV \citep{prinja90} velocities
of 1520 km\,s$^{-1}$, 1450 km\,s$^{-1}$ and 1460 km\,s$^{-1}$, respectively, the latter obtained from the blue absorption edge of the UV P Cygni  C{\sc iv}\,1550\AA\ profile.


\begin{figure}
\centering
\includegraphics[width=0.35\textwidth,bb=60 30 552 796]{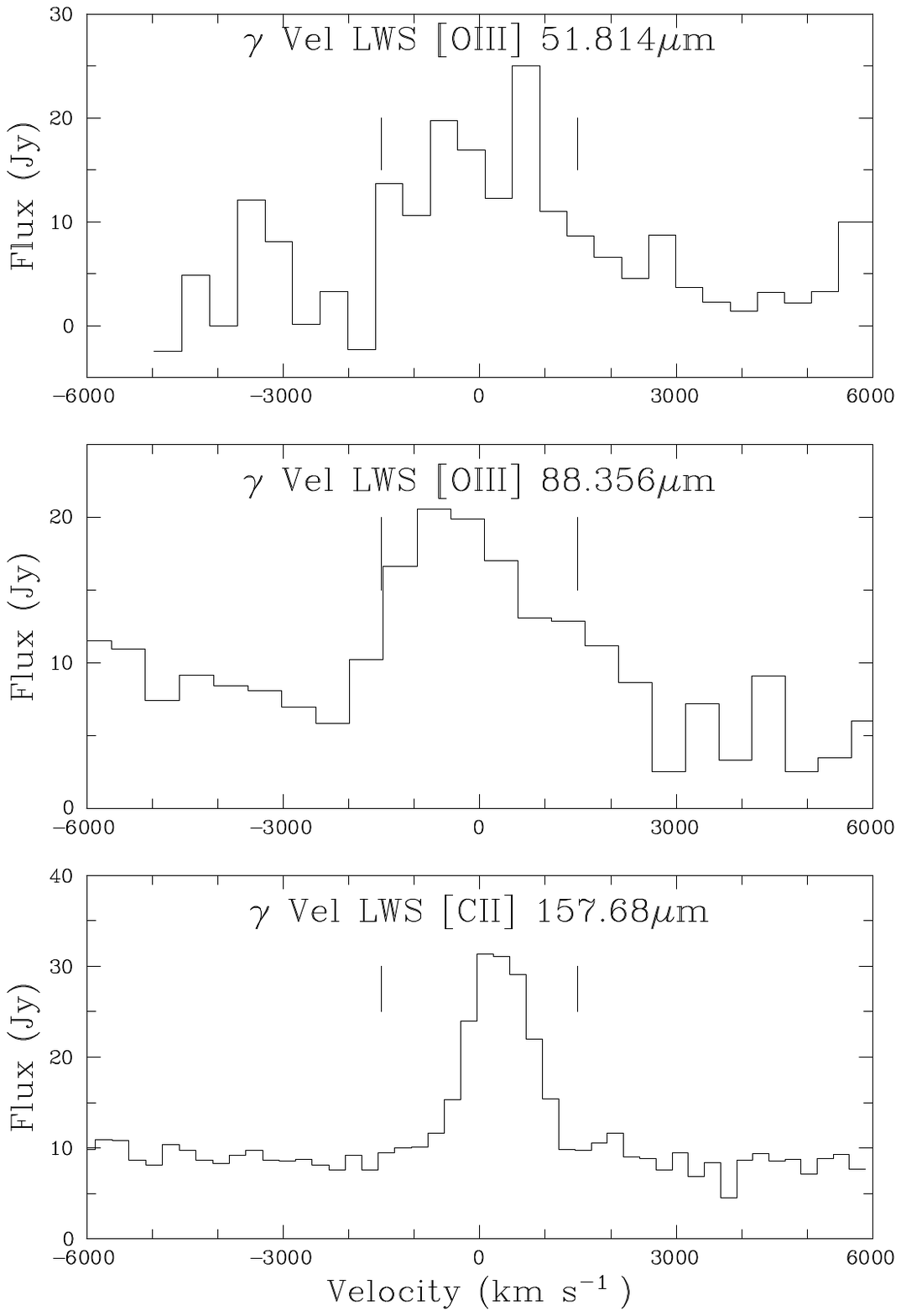}
\caption{{\it ISO}-LWS spectrum of $\gamma$ Vel, an average of back and forth scans, plotted in velocity space 
about a central wavelength [O\,{\sc iii}] $51.81\mu$m (top), [O\,{\sc iii} 88.36$\mu$m (centre) and [C\,{\sc ii}] 157.68$\mu$m (bottom). Vertical lines indicate the expected 
extent of the stellar wind $v_\infty\,{=}\,$1500 km\,s$^{-1}$. }
\label{fig:isowr11}
\end{figure}

\subsection{Archival UV/optical spectroscopy}


Extensive archival UV/optical spectroscopy of $\gamma$ Vel exists, including  {\it Copernicus} \citep{Johnson78}, {\it IUE} \citep{stlouis93} and ESO 0.5m HEROS \citep{Schmutz97}. {\it Copernicus} far-UV and near-UV
spectroscopy covers $\lambda\lambda$946--3175 at a resolution of 0.2--0.4\AA, while high resolution {\it IUE} spectroscopy with the SWP and LWR cameras cover
1150--3300\AA. We have utilised calibrated {\it IUE} observations at phase 0.5 \citep{stlouis93}, plus far-UV {\it Copernicus} U2 spectroscopy from March 1977, although no official flux calibration exists for the latter. The H\,{\sc i} column density towards $\gamma$ Vel is $\log N ({\rm H}$ {\sc i}) = 19.8  cm$^{-2}$ according to \citet{1975ApJ...200..402B} and \citet{1976ApJ...203..378Y}.

In the optical, \citet{Aller64} discuss the optical morphology of $\gamma$ Vel. We use phase-averaged ESO 0.5m HEROS spectroscopy covering 0.35--0.55$\mu$m and 0.58--0.86$\mu$m \citep{Schmutz97, demarco99}. Since these were not flux calibrated, we utilise relative intensities of optical emission lines from \citet{Aller64} to produce a relatively calibrated optical spectrum for the 0.4--0.55$\mu$m HEROS dataset, which is put on an absolute scale via narrow band magnitudes $b$=1.42 and $v$=1.70 \citep{demarco00}. Violet and red HEROS dataset is anchored to the composite continua of the O+WC8 model. We prefer narrow-band filter optical magnitudes to standard broad-band photometry \citep{Johnson66} owing to emission line contamination of the latter. Since HEROS coverage excludes the C\,{\sc iii} $\lambda$5696 and C\,{\sc iv} $\lambda$5808 classification lines, we note  $F_{\rm 5696}/F_{\rm 5808}$ = 1.4 according to \citet{Aller64}.

Since $\gamma$ Vel is a binary system, we adopt a flux ratio of $F_{v}$(O)/$F_{v}$(WC8) = 3.61 following \citet{demarco00}, i.e. $M_{v}$(WC8) --$M_{v}$(O) = +1.39 mag. Using our adopted distance modulus (7.63 mag) and extinction of E$_{\rm B-V}$ = 0.09 mag imply a systemic absolute magnitude of $M_{v}$ = --6.24 mag, with $M_{v}$(O) = --5.97 mag and $M_{v}$(WC8) = --4.58 mag. The relative contribution of its constituents will vary with wavelength, with O to WR continuum light ratios of 3.7 at 0.5$\mu$m and 1.0 at 2.0$\mu$m \citep{demarco99}.

\begin{figure*}
\centering
\begin{minipage}[c]{0.75\linewidth}
\includegraphics[width=0.65\textwidth,angle=-90,bb=30 30 535 784]{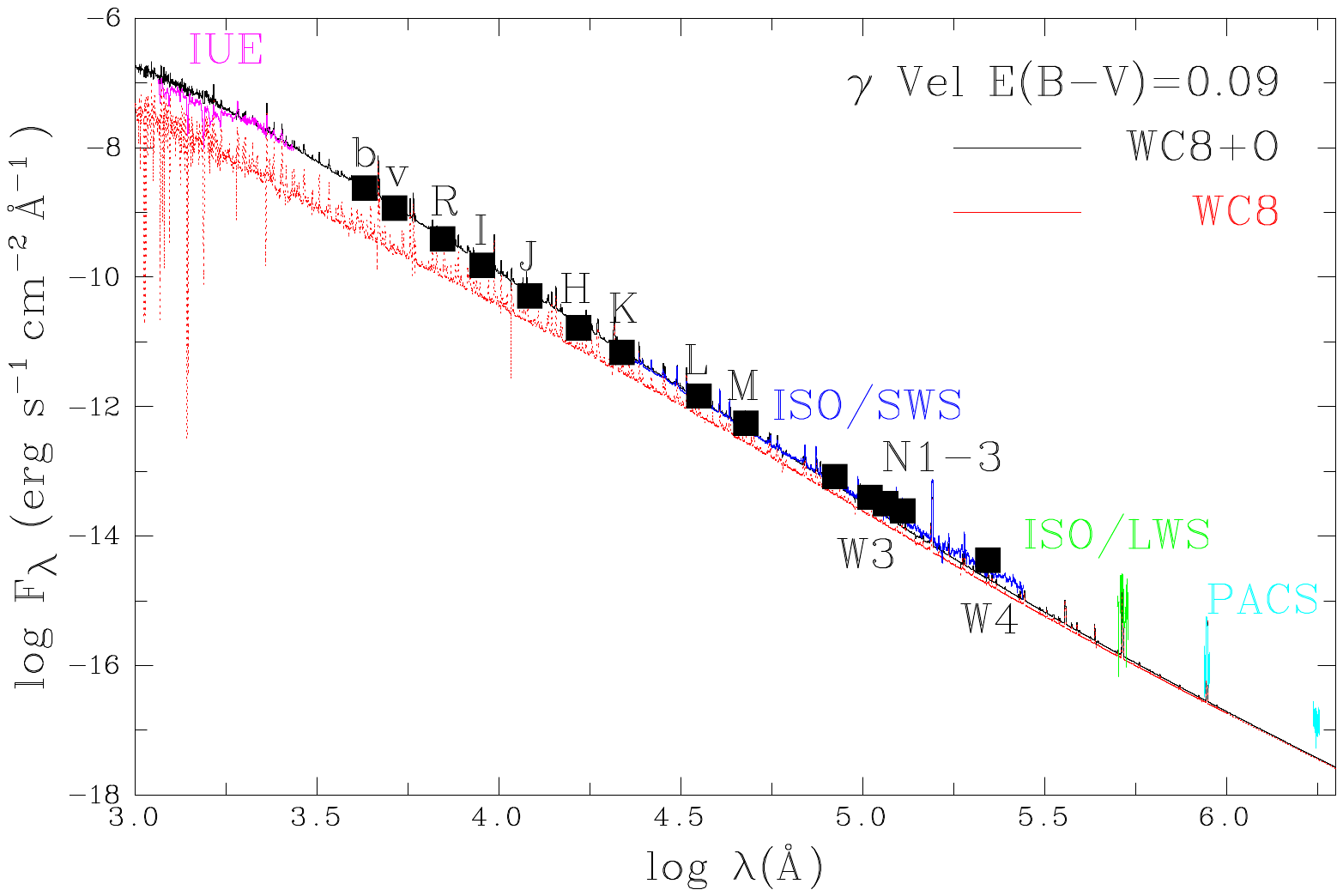}
\end{minipage}\hfill
\begin{minipage}[c]{0.25\linewidth}
\centering
  \caption{De-reddened spectral energy distribution of $\gamma$ Vel from {\it IUE} (pink),  {\it ISO} SWS (blue) and LWS ([O\,{\sc iii}] 52$\mu$m region only, green) and {\it Herschel} PACS spectrophometry (cyan), together with combined theoretical WC8+O model (black) and WC8 (red) model, plus optical-IR photometry \citep{Johnson66,Williams90,demarco00,WISE}.} 
	\label{fig:wr11}
	\end{minipage}
\end{figure*}

\subsection{Archival IR spectroscopy}

 \citet{Barnes74} and \citet{Aitken82} present near-IR spectroscopy of $\gamma$ Vel, supplemented by He\,{\sc i} 1.083$\mu$m spectroscopy from ESO's New Technology Telescope (NTT)+EMMI  \citep{demarco00}.  IR photometry between 1--13$\mu$m is obtained from average measurements by \citet{Williams90}, supplemented by bands 3--4 from {\it WISE} \citep{WISE}. {\it ISO} SWS \citep{deGraauw96} mid-IR spectroscopy covering 2.4--27.5$\mu$m ($R$ = 820--1700) includes a mix of inner wind and fine-structure lines and has been discussed by \citet{vanderHucht96}. {\it ISO} LWS \citep{Clegg96} far-IR spectroscopy (PI M.J. Barlow) obtained from the {\it ISO} archive\footnote{http://iso.esac.esa.int/ida/}, cover 43--197$\mu$m. $\gamma$ Vel suffers from low S/N  in the long wavelength SWS (30--45$\mu$m) and short wavelength LWS  (43--90$\mu$m) channels of {\it ISO}. Nevertheless we have been able to use the average fluxed spectra of back and forth scans to identify several fine structure lines [O\,{\sc iii}] 51.81$\mu$m, 88.36$\mu$m and [C\,{\sc ii}] 157.7$\mu$m,  as shown in Fig.~\ref{fig:isowr11}. 



The {\it ISO} [O{\sc iii}]$\,51.81\micron$ spectrum of $\gamma$ Vel is very noisy, posing a problem for the measurement of
line flux as the continuum level is difficult to determine. We measure the flux
of this line in two ways; firstly using the \texttt{FLUX} command from the Starlink  spectroscopic package {\sc dipso} \citep{dipso} 
between the expected velocity limits after
subtraction of a by-eye estimate of the continuum, and secondly by fitting a
Gaussian profile to the line using {\sc dipso}'s Emission Line Fitting (ELF) suite, after
subtraction of a (2nd order) polynomial fit to the continuum. Although the
emission line is expected to be flat-topped and highly non-Gaussian, this
structure is lost in the noise, permitting us to employ Gaussian fitting as an
approximation.  From the \texttt{FLUX} method, multiple measurements give an
average [O\,{\sc iii}] 51.81$\mu$m line flux of
$(8.1\pm0.6) {\times}\,10^{-12}~\mathrm{erg}\,\mathrm{s}^{-1}\mathrm{cm}^{-2}$,
and the Gaussian fit gives a compatible
$(7.8\pm1.6) {\times}\,10^{-12}~\mathrm{erg}\,\mathrm{s}^{-1}\mathrm{cm}^{-2}$. We
adopt a line flux of
$(8\pm1)\,{\times}\,10^{-12}~\mathrm{erg}\,\mathrm{s}^{-1}\mathrm{cm}^{-2}$, incorporating 10\% flux calibration uncertainties for `medium' point sources \citep{2000ESASP.456..267G}. 

The 84--93$\mu$m region is covered by the first order of LW1 detector
of {\it ISO} LWS, as well as second order of the SW5 detector. The standard data product adopts the higher resolution SW5 spectrum, but LW1 has significantly
higher sensitivity permitting [O\,{\sc iii}] 88.36$\mu$m to be detected (Fig.~\ref{fig:isowr11}). The integrated flux measured using {\sc dipso's} ELF package is $(4.2\pm0.4) {\times}\,10^{-12}~\mathrm{erg}\,\mathrm{s}^{-1}\mathrm{cm}^{-2}$,
20\% higher than the {\it Herschel} PACS measurement. We favour the latter in our analysis owing to its improved S/N and order of magnitude improved spectral resolution. 
Finally, [C\,{\sc ii}] 157.68$\mu$m is prominent in the long wavelength LWS channel. This line has a flux of $(1.84\pm0.08) {\times}\,10^{-12}~\mathrm{erg}\,\mathrm{s}^{-1}\mathrm{cm}^{-2}$, although it is spectrally
unresolved since its FWHM matches the {\it ISO}-LWS grating resolution of $R\sim$260, so does not arise in the stellar wind (its critical
density  is $\sim 50$ cm$^{-3}$).


\section{Spectroscopic analysis}\label{cmfgen}

The detailed spectral analysis of the O and WC components in $\gamma$ Vel by \citet{demarco99} and \citet{demarco00} form the basis of our 
study. We utilise the non-LTE line-blanketed model atmosphere code {\sc cmfgen} \citep{hillier98} which solves the radiative transfer equation in the co-moving frame, subject to statistical
and radiative equilibrium, assuming an expanding, spherically-symmetric, homogenous or clumped, time independent atmosphere. Line blanketing is treated correctly in the transfer problem except that a `super level' approach is used, involving the combining of levels with similar energies and properties into a single super level.

\begin{table*}
\centering
\caption{Physical, wind and chemical parameters for the WC8 and O7.5\,III components of $\gamma$ Vel from {\sc cmfgen} analysis, updated from \citet{demarco99} and \citet{demarco00}. These include radially-dependent volume filling factors, $f_{\log N_e}$, for the WC8 component at various electron densities ($N_{e}$/cm$^{-3}$), corresponding to the line forming regions of optical lines, mid-IR (e.g. [Ne\,{\sc iii}] 15.5$\mu$m) and far-IR (e.g. [O\,{\sc iii}] 88$\mu$m) fine structure lines, respectively.}
\centering
\begin{tabular}{ l @{\hspace{2mm}} c @{\hspace{2mm}} c @{\hspace{2mm}} c @{\hspace{2mm}}c @{\hspace{2mm}} c @{\hspace{2mm}} c @{\hspace{2mm}} c @{\hspace{2mm}} c @{\hspace{2mm}} c @{\hspace{2mm}} c @{\hspace{2mm}} c @{\hspace{2mm}} c @{\hspace{2mm}} c @{\hspace{2mm}} c @{\hspace{2mm}} c @{\hspace{2mm}} c @{\hspace{2mm}} c @{\hspace{2mm}} c @{\hspace{2mm}} c @{\hspace{2mm}} c }
\hline
Star & $T_\ast$ & $T_{\rm eff}$ & $R_\ast$ & $\log g$ & $ \log L$ &  $v_{\infty}$  & log($\dot{M}$)       & $R_{\rm max}$  & $f_{\rm 12}$ & $f_{\rm 5}$ & $f_{\rm 0}$ &  $M_{\rm v}$ & $X_{\rm H}$ & $X_{\rm He}$ & $X_{\rm C}$ & $X_{\rm N}$ & $X_{\rm O}$  & $X_{\rm Ne}$ & $X_{\rm S}$ & $X_{\rm Fe}$ \\
  &  kK &   kK & $R_{\odot}$ &    cm\,s$^{-2}$      & $L_{\odot}$              & km\,s$^{-1}$ & $M_\odot$ yr$^{-1}$ & $R_{\ast}$ & &  & & mag & \% & \% & \% & \% & \% & \%  & \% \\
\hline
WC8      & 90     &  47.5 & \phantom{0}1.9 & 4.85 & 5.31 & 1420 & $-4.84 $  & 4.5$\times 10^{6}$  & 0.1      & 0.5                & 0.5 & --4.6 & 0  & 67  & 30 & 0     & 1.1 & 1.7   & 0.04 & 0.1 \\ 
O7.5\,III & 35.1  &  35.0  & 16.2                  & 3.47 & 5.56 & 2500 & $-6.55$   & 2.0$\times 10^{2}$  & \multicolumn{3}{c}{ --- 1.0 --- } & --6.0 & 70 & 28 & 0.3 & 0.1 & 1 & 0.2 & 0.04 & 0.1 \\ 
\hline
\end{tabular}\par
\label{tab:params}
\end{table*}

\begin{figure*}
\centering
\includegraphics[width=0.315\textwidth,angle=-90,bb=30 30 535 784]{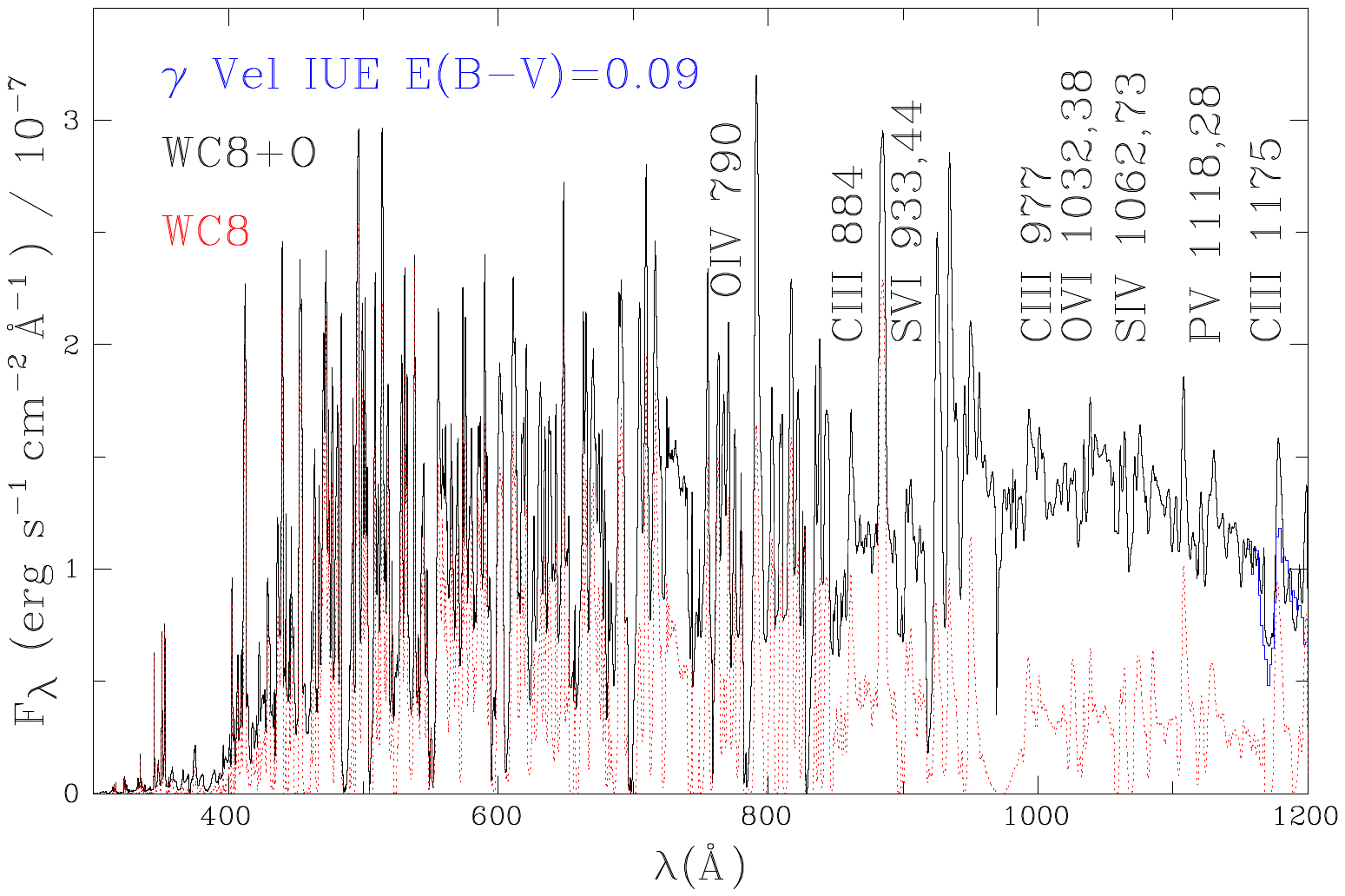}
\includegraphics[width=0.315\textwidth,angle=-90,bb=30 30 535 784]{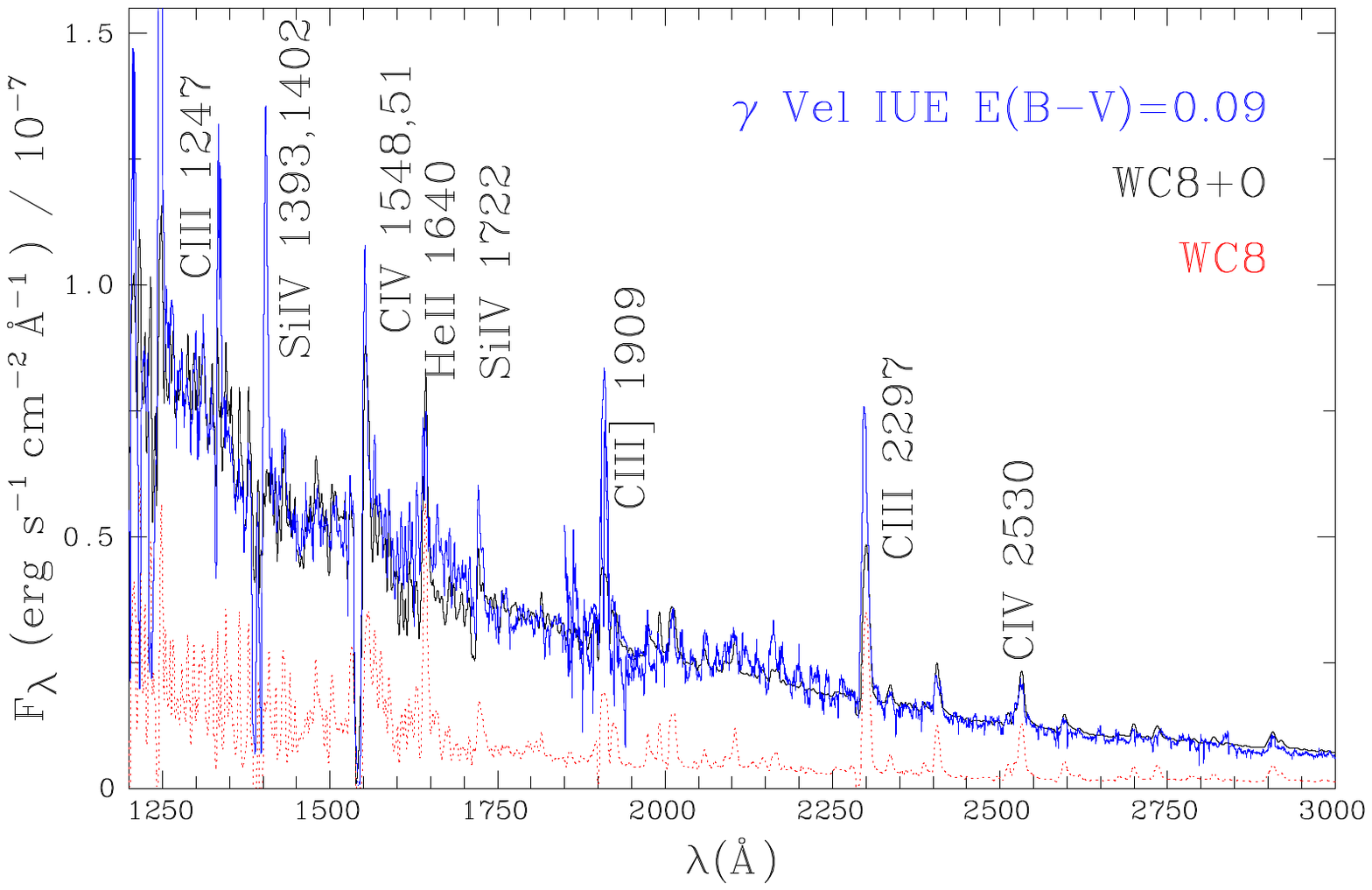}
\includegraphics[width=0.315\textwidth,angle=-90,bb=30 30 535 784]{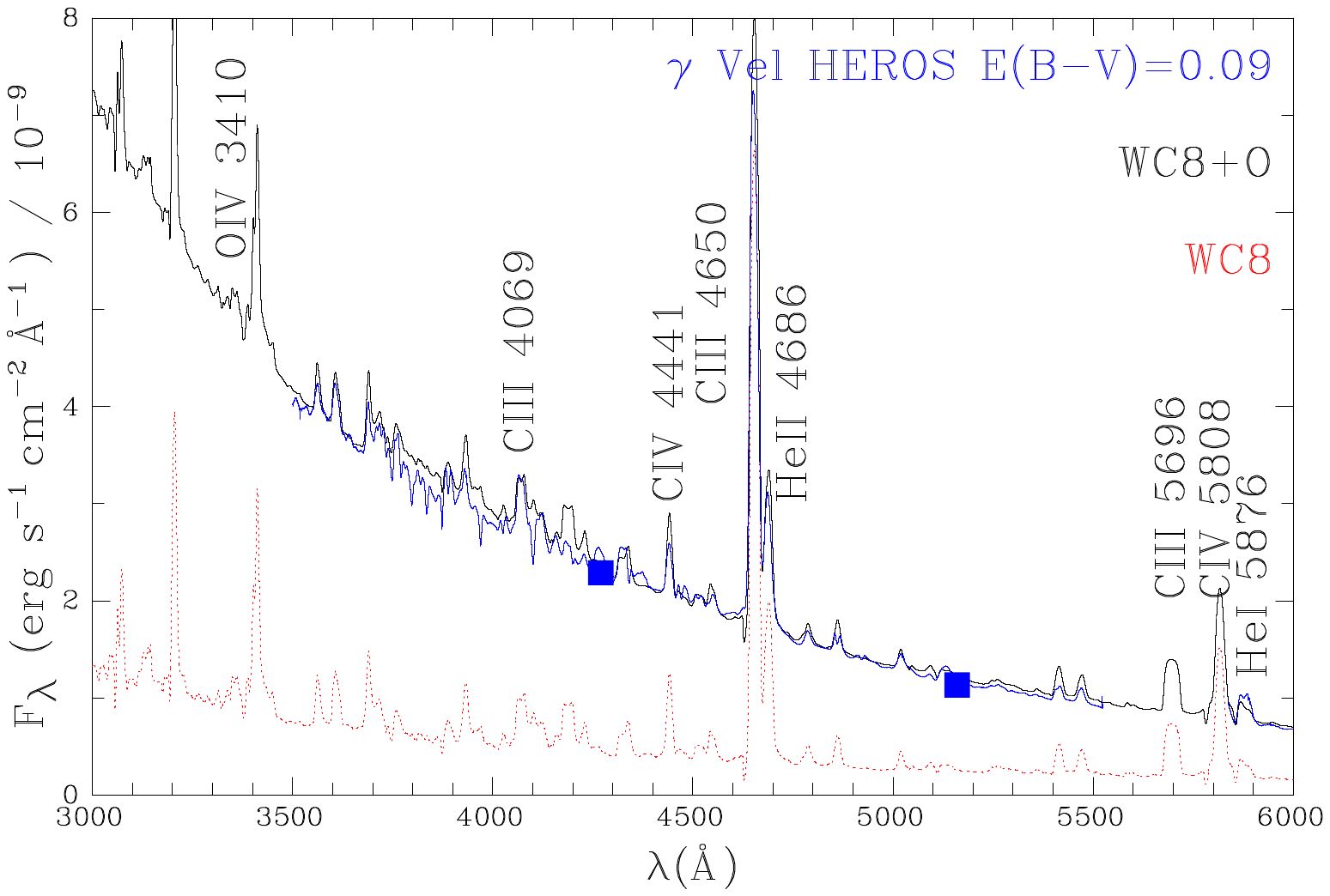}
\includegraphics[width=0.315\textwidth,angle=-90,bb=30 30 535 784]{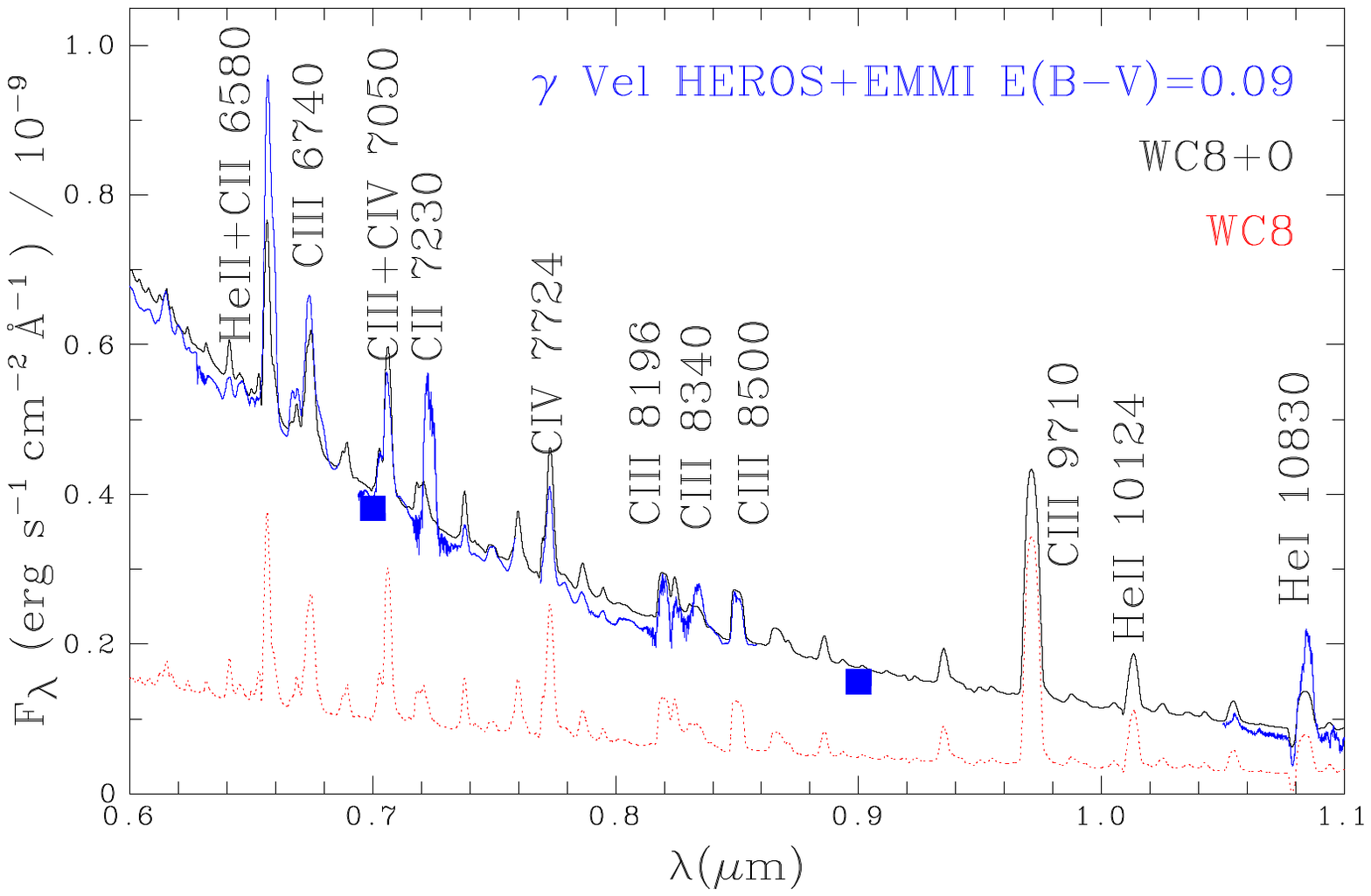}
\includegraphics[width=0.315\textwidth,angle=-90,bb=30 30 535 784]{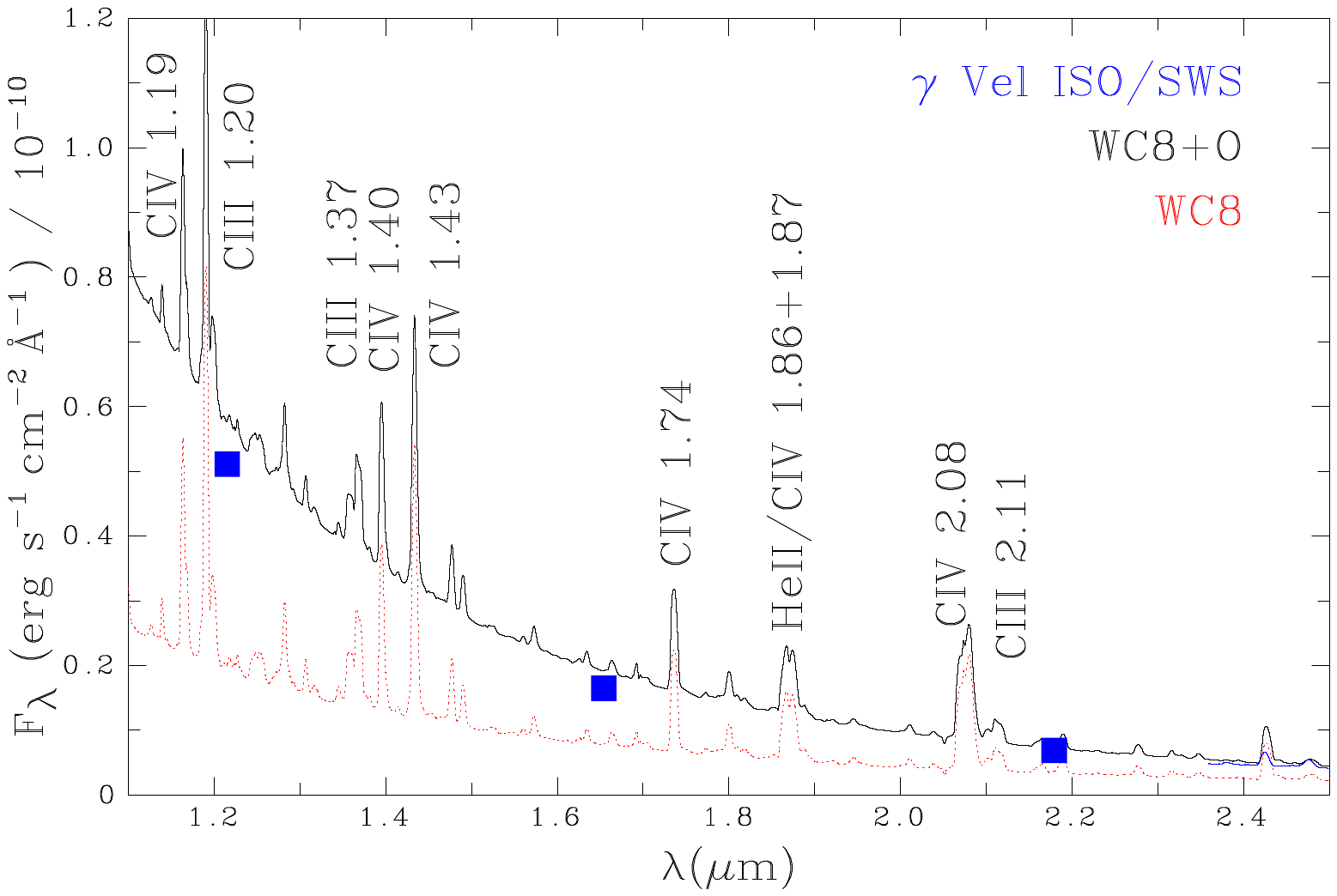}
\includegraphics[width=0.315\textwidth,angle=-90,bb=30 30 535 784]{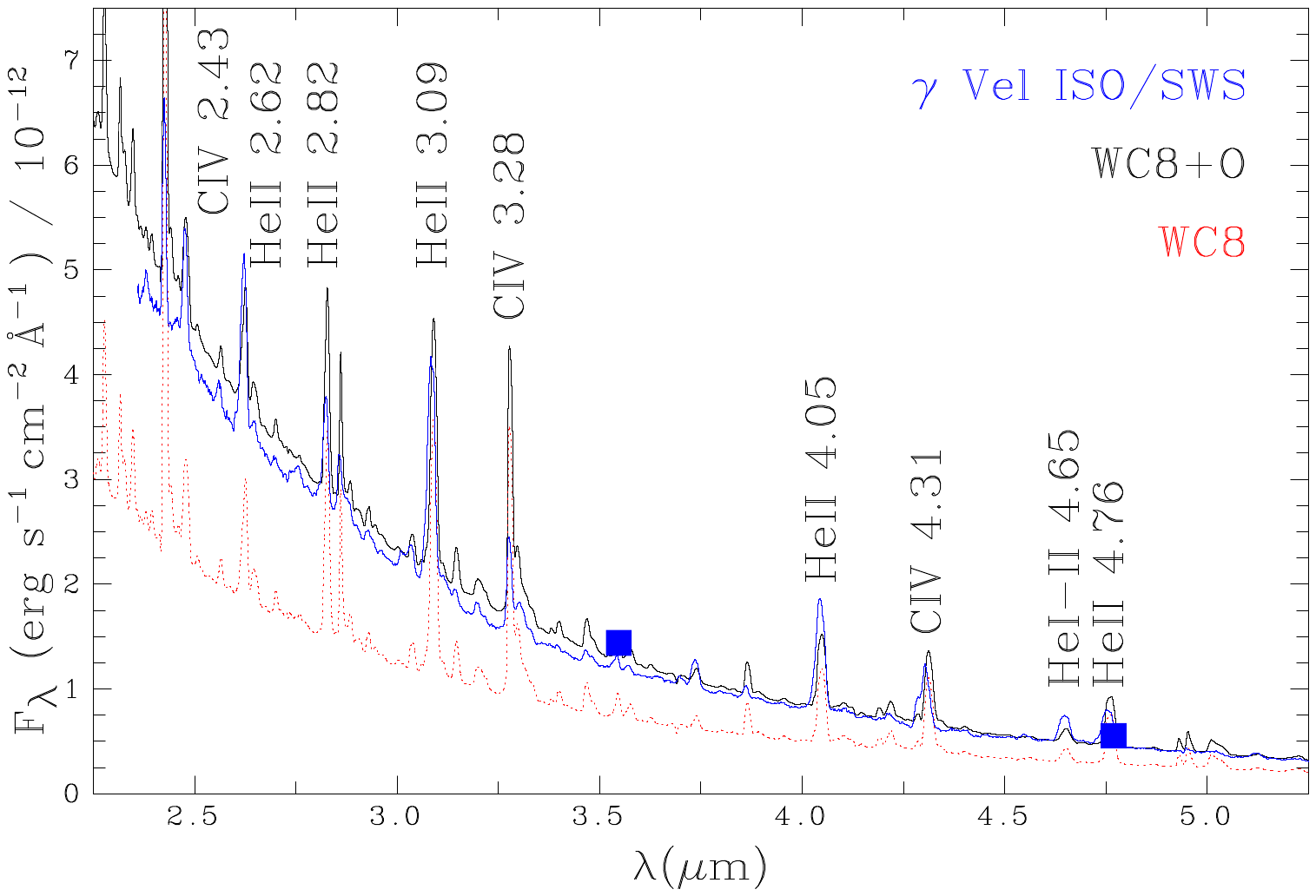}
\includegraphics[width=0.315\textwidth,angle=-90,bb=30 30 535 784]{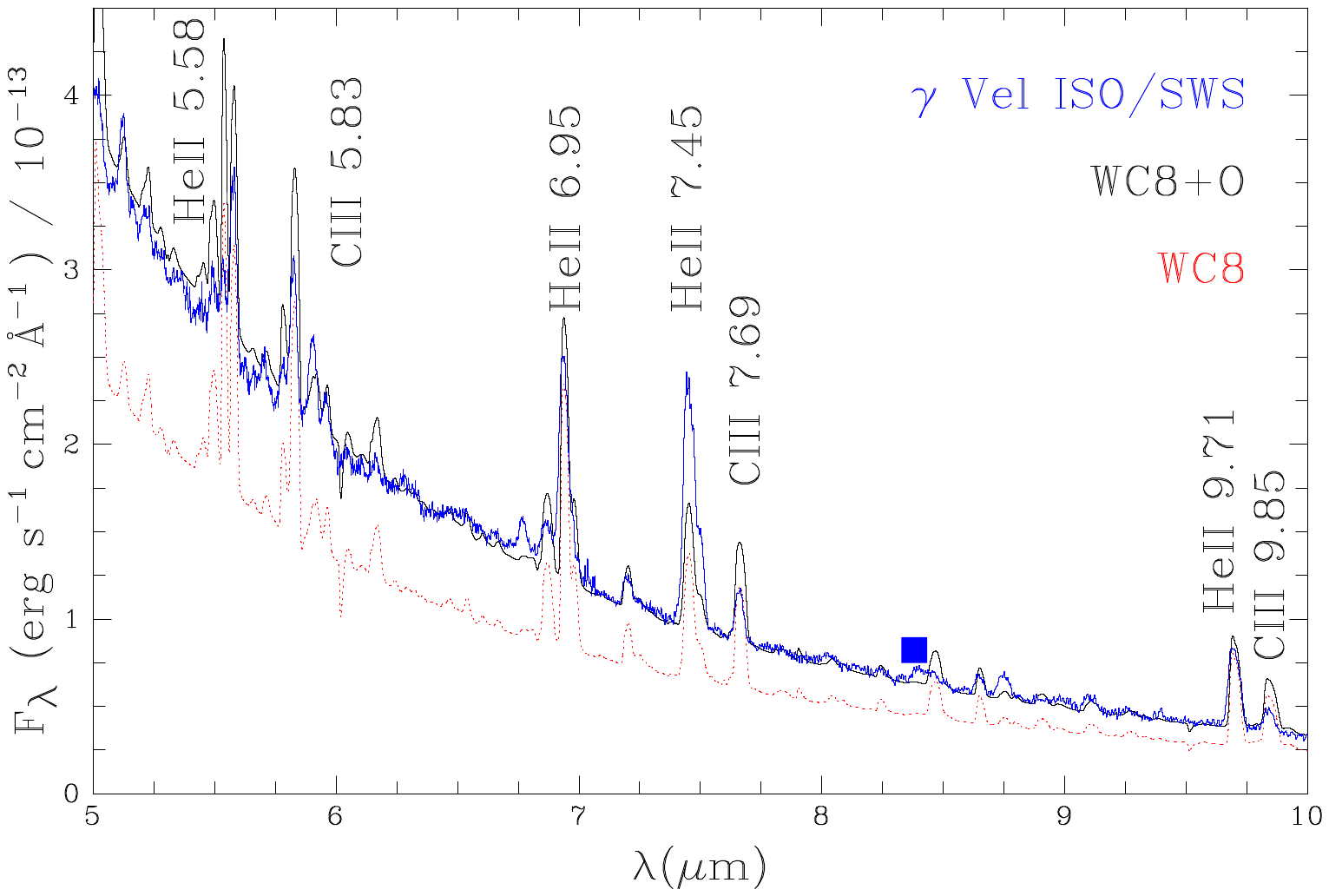}
\includegraphics[width=0.315\textwidth,angle=-90,bb=30 30 535 784]{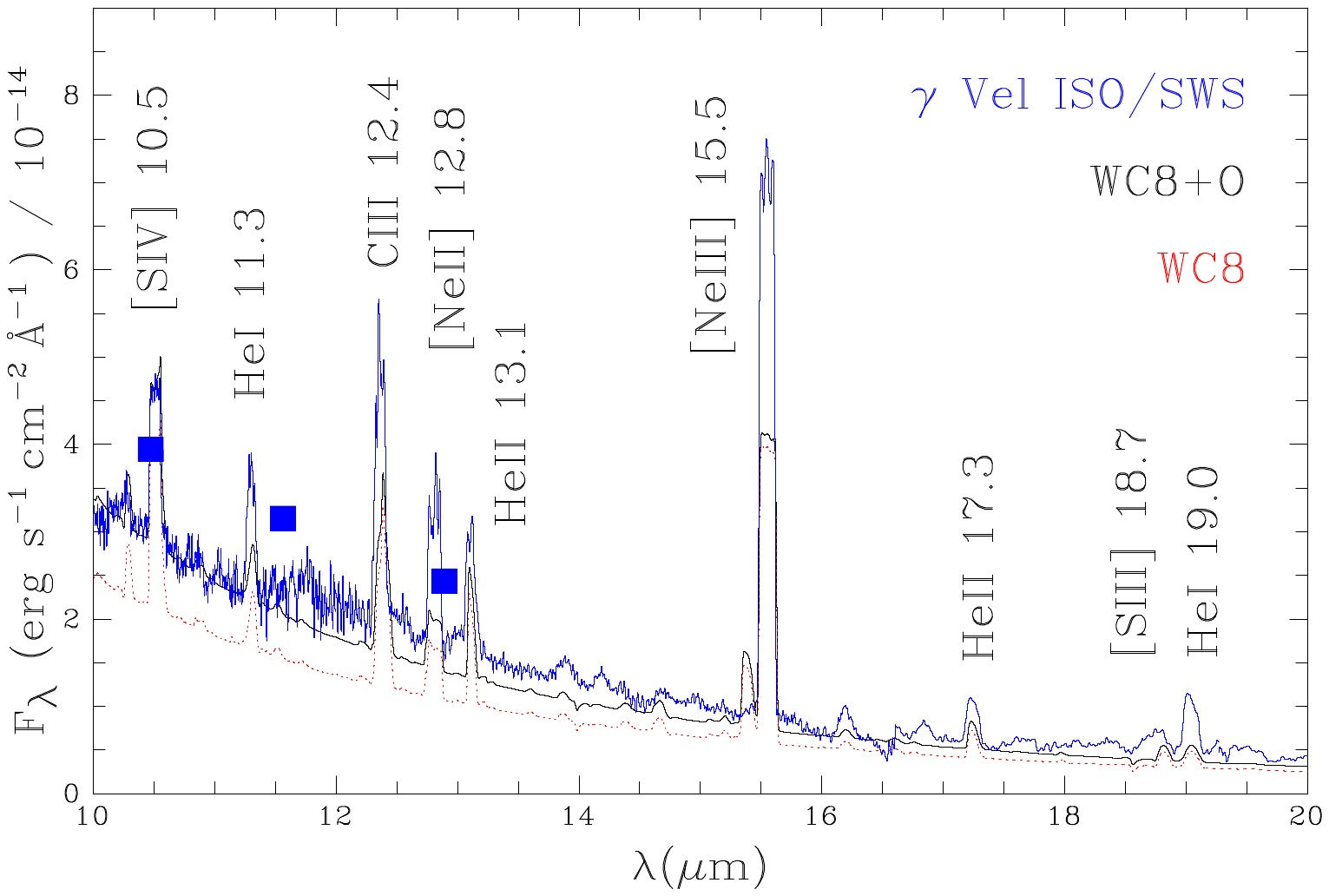}
\caption{De-reddened UV {\it IUE/HIRES} spectroscopy (phase 0.5), calibrated optical ESO 50cm+HEROS and NTT+EMMI He\,{\sc i} $\lambda$10830 spectroscopy, {\it ISO} SWS infrared spectroscopy of $\gamma$ Vel (blue), together with theoretical WC8+O model (black) and WC8 component (red), excluding correction for interstellar H\,{\sc i}. Optical-IR photometry \citep{Johnson66,Williams90,demarco00,WISE} is presented as blue boxes.}
\label{fig:spectrum}
\end{figure*}


\subsection{O star}

For the O star component of $\gamma$ Vel we include the following elements: H, He, C, N, O, Ne, Na, Mg, Al, Si, P, S, Ca, Fe, Ni, with standard Milky Way abundances. We adjust the physical parameters obtained by \citet{demarco99} to the modern
distance of 336 pc, but otherwise do not attempt to revise its solution, since our focus is on the outer wind of the WC star. Nevertheless, we note that the composite WC+O HEROS spectroscopy favours a high rotation rate of $v_{\rm eq} \sin i \sim$ 350 km\,s$^{-1}$ for the O star from comparison with He\,{\sc i} $\lambda$4471 \citep[][their fig.~5]{Schmutz97}, somewhat higher than the previous estimate of $v_{\rm eq} \sin i \sim$ 220 km\,s$^{-1}$ \citep{1990A&A...240..105B}. High rotation velocities are expected for the mass gainer in a post-mass transfer system  \citep{deMink13}.
We use a hydrostatic solution at depth, together with a standard $\beta$ = 1 law for the wind, and adopt a smooth wind (the IR energy distribution of $\gamma$ Vel is dominated by the WC component). The outer boundary of the O star
model is set at 200 $R_{\ast}$, corresponding to $N_e = 5 \times 10^{4}$ cm$^{-3}$ and $T_{e}$ = 9500 K. Physical and wind properties are summarised in Table~\ref{tab:params}.

\begin{figure*}
\centering
\includegraphics[width=0.4\textwidth,bb=40 45 720 585]{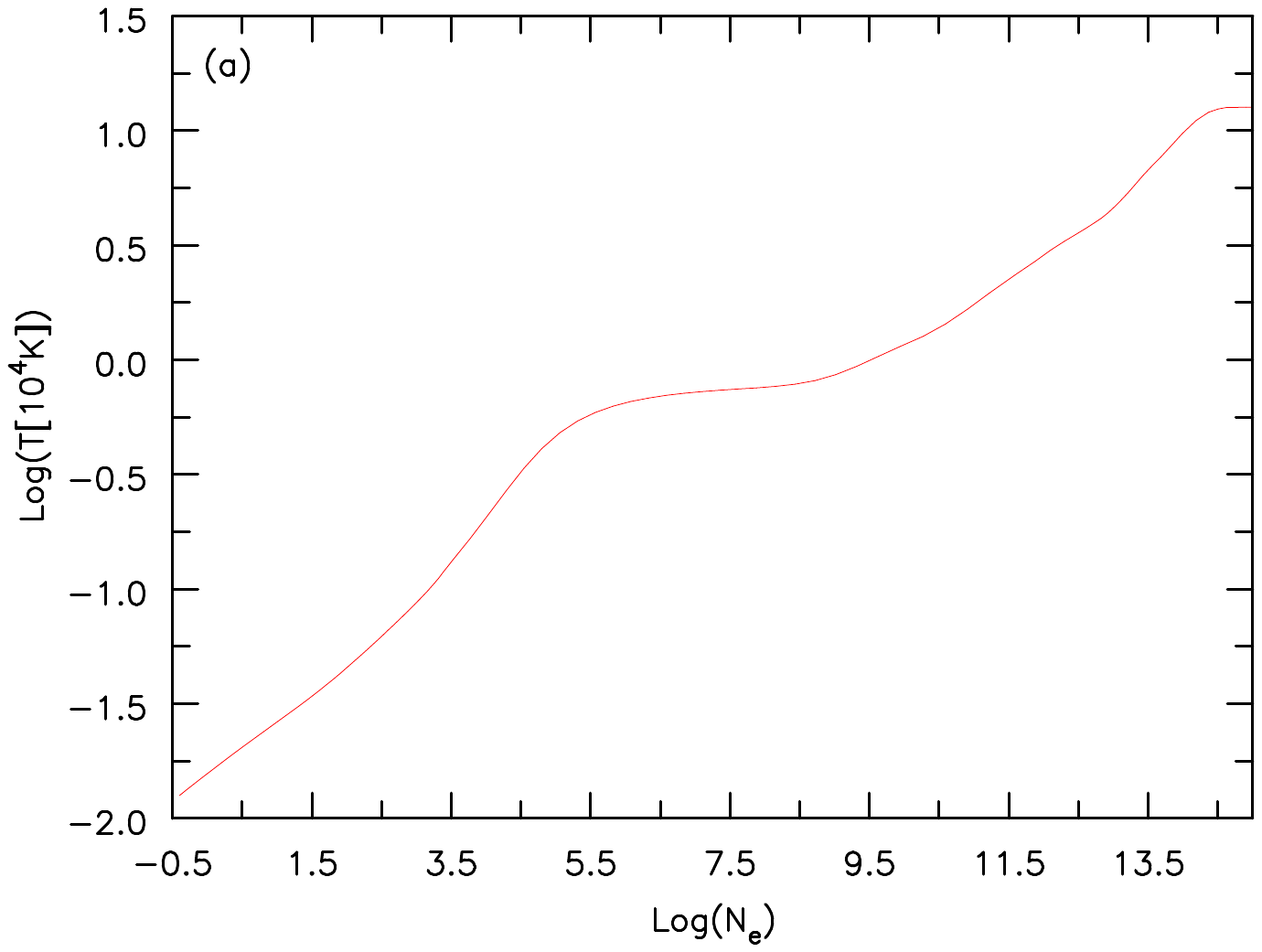}
\includegraphics[width=0.4\textwidth,bb=40 45 720 585]{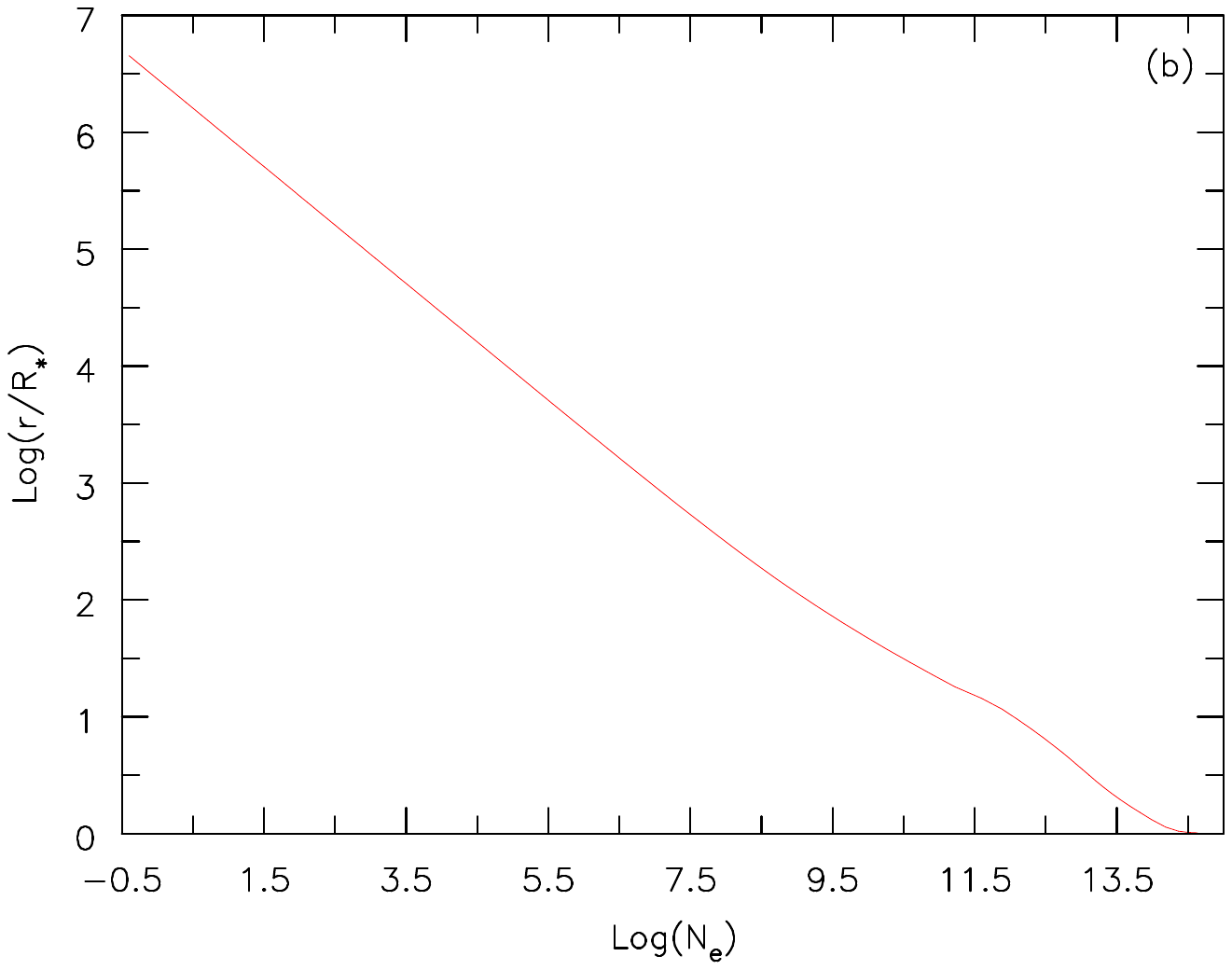}
\includegraphics[width=0.4\textwidth,bb=40 45 720 585]{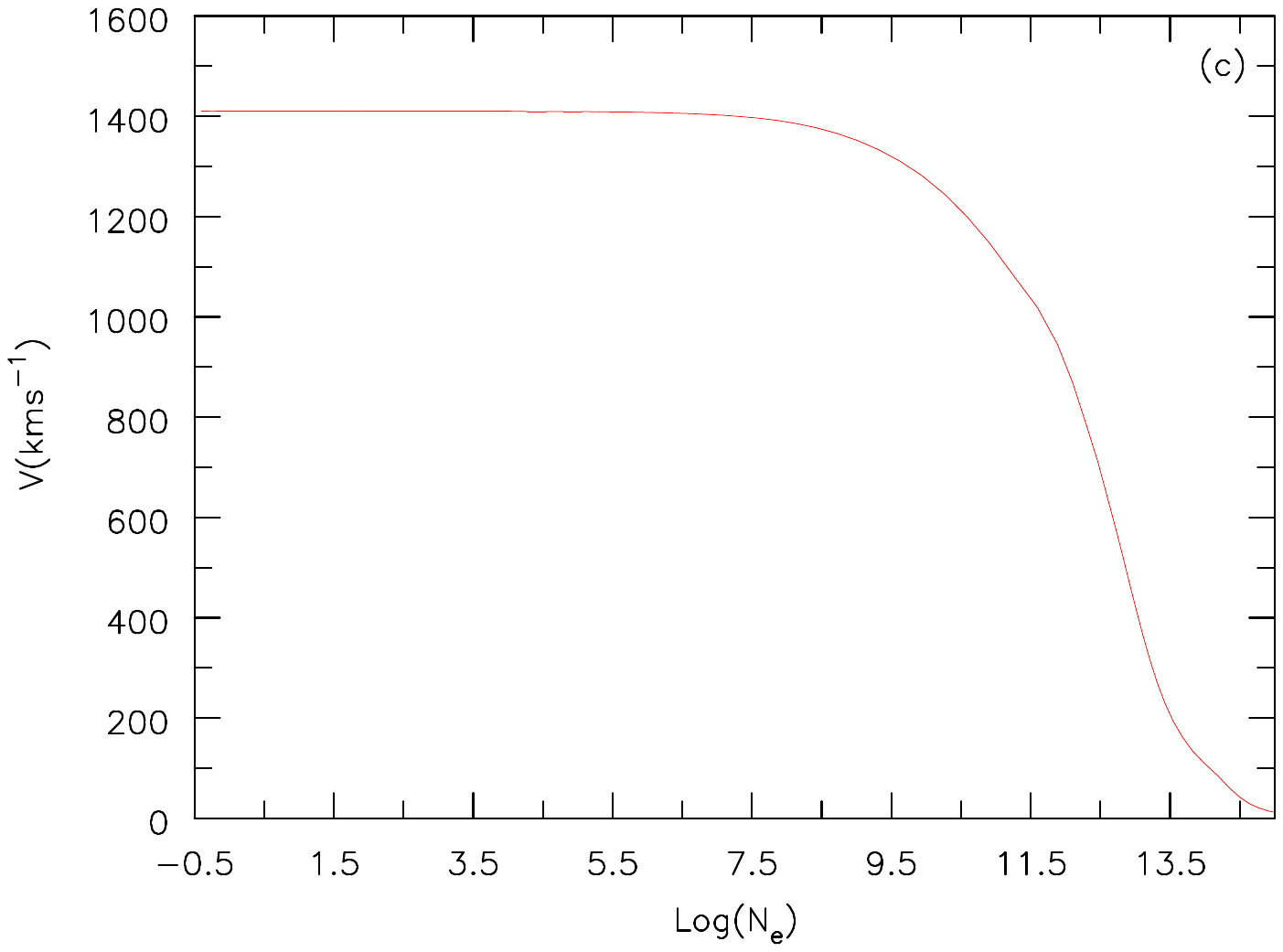}
\includegraphics[width=0.4\textwidth,bb=40 45 720 585]{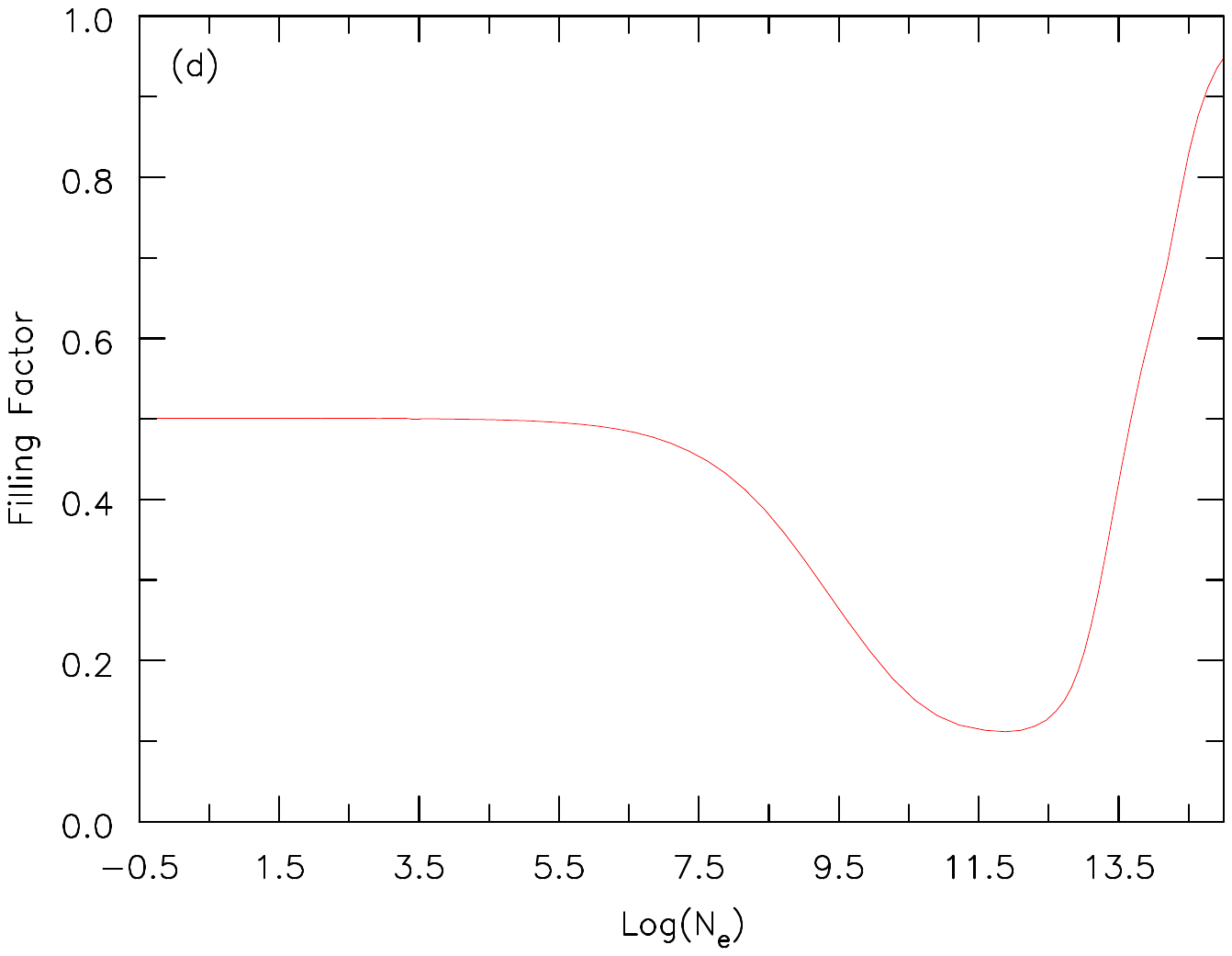}
\includegraphics[width=0.4\textwidth,bb=40  45 720 585]{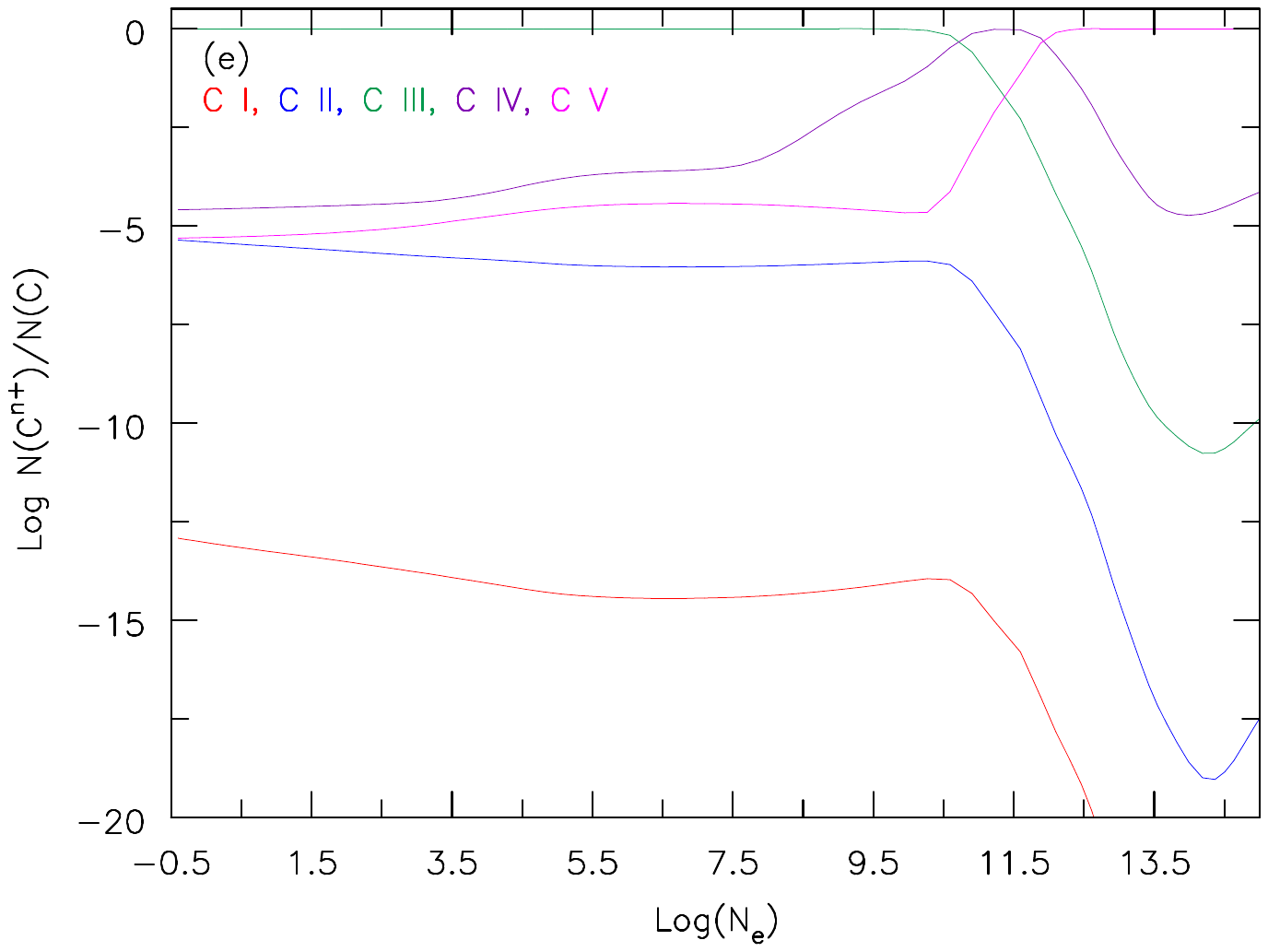}
\includegraphics[width=0.4\textwidth,bb=40  45 720 585]{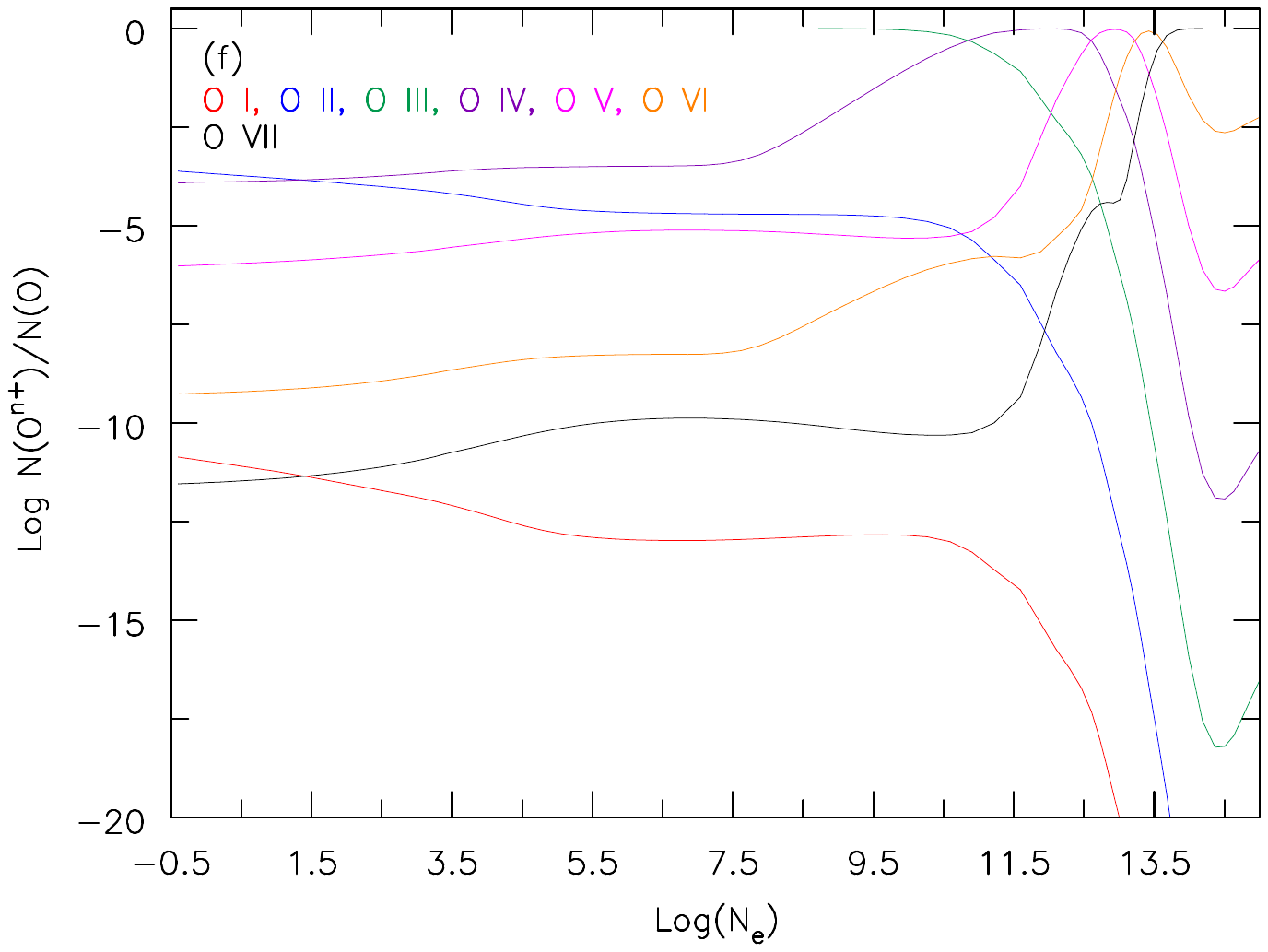}
\includegraphics[width=0.4\textwidth,bb=40  45 720 585]{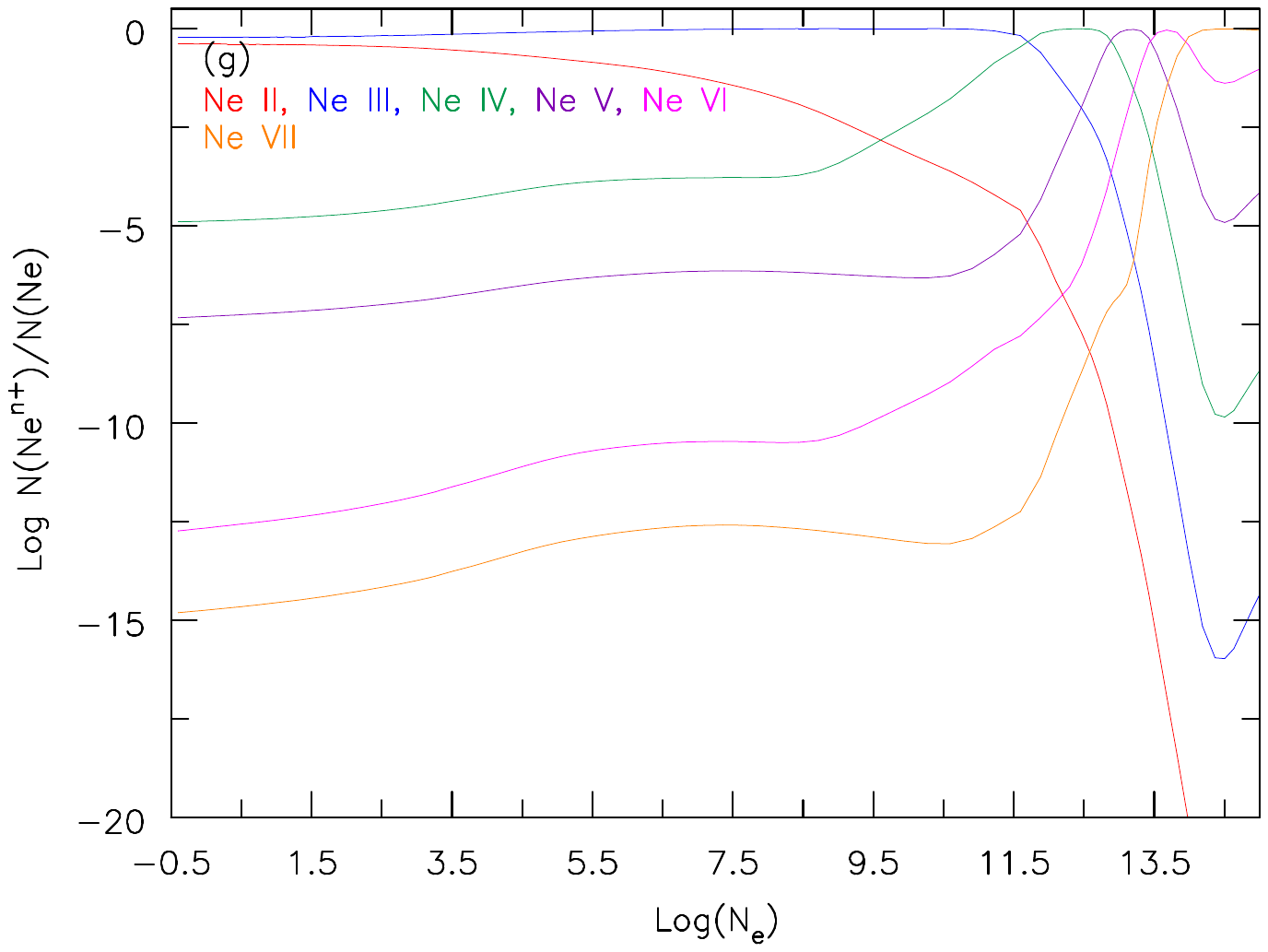}
\includegraphics[width=0.4\textwidth,bb=40  45 720 585]{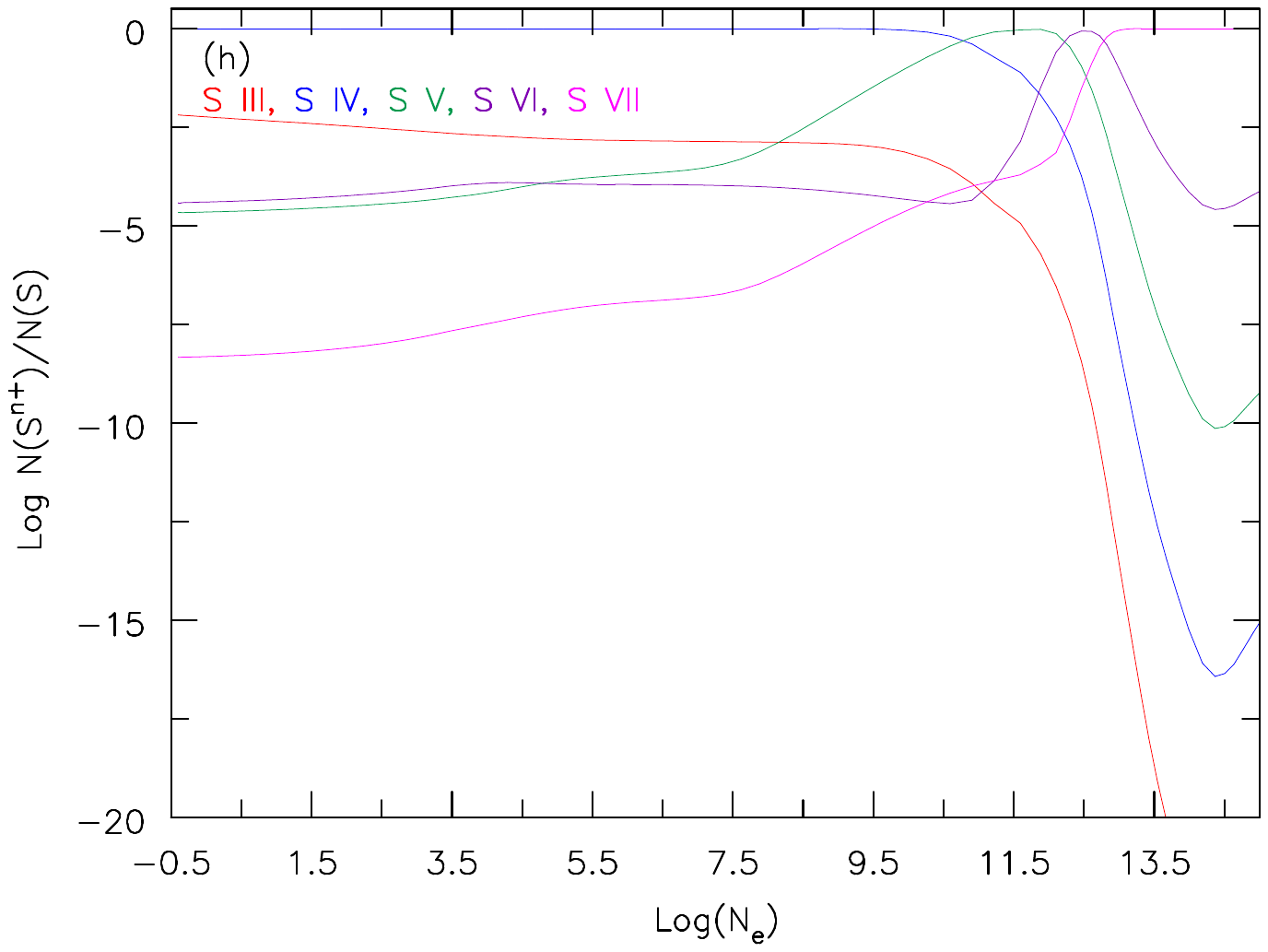}
\caption{WC8 wind structure for $\log N_{e}$ (cm$^{-3}$) versus (a) $\log T_{e}$  (10$^{4}$ K, red); (b)  $\log r/R_{\ast}$ (red); (c) velocity $v(r)$ (km\,s$^{-1}$, red);  (d) clumping volume filling factor $f$ (red); ionization structure of (e) carbon: $\log$ C$^{0}$/C (red), C$^{+}$/C (blue), C$^{2+}$/C (green), C$^{3+}$/C (purple), C$^{4+}$/C (pink); (f) oxygen: $\log$ O$^{0}$/O (red), O$^{+}$/O (blue), O$^{2+}$/O (green), O$^{3+}$/O (purple), O$^{4+}$/O (pink), O$^{5+}$/O (orange), O$^{6+}$/O (black); (g) neon: $\log$ Ne$^{+}$/Ne (red), Ne$^{2+}$/Ne (blue), Ne$^{3+}$/Ne (green), Ne$^{4+}$/Ne (purple), Ne$^{5+}$/Ne (pink), Ne$^{6+}$/Ne (orange); (h) sulphur: $\log$ S$^{2+}$/S (red), S$^{3+}$/S (blue), S$^{4+}$/S (green), S$^{5+}$/S (purple), S$^{6+}$/S (pink). Higher ionization stages will dominate optically thick regions at high densities ($\log N_{e} \geq 13.5$).}
\label{fig:wind_struc}
\end{figure*}

\subsection{WC8 star}

For the WC component of $\gamma$ Vel we include the following elements: He, C, O, Ne, Si, S, Ar, Ca, Fe, Ni. Intermediate mass elements (Ar, Ca) are included since they have a strong influence on the extreme UV blanketing (complementing Fe, Ni) which affects the strength of key optical diagnostics including C\,{\sc iii} $\lambda$5696, such that higher temperatures are necessary to reproduce optical diagnostic lines (C\,{\sc iii} $\lambda$5696, C\,{\sc iv} $\lambda\lambda$5801-12) with respect to previous studies \citep{demarco00}. 

For the velocity law, we use a hydrostatic solution at depth, together with a standard $ \beta$ = 1 for the outer wind. We adopt a turbulent velocity of 100 km\,s$^{-1}$, requiring $v_{\infty}$ = 1420 km\,s$^{-1}$ in order to reproduce the mid-IR fine structure lines. The emerging consensus is that clumping persists to large radii
(at least $10^3\,R_\ast$) in a line-driven wind, characterised by a clumping (volume filling) factor which disperses to a minimum of $4\mbox{--}5$
($f{=}\,0.2\mbox{--}0.25$) according to \citet{runacres05, RubioDiez22}. Consequently we adopt a non-standard radially dependent volume filling factor $f(r)$ of the form 
\[
f(r) = f_{\alpha} + (1 - f_{\alpha}) \exp (-v(r)/f_{\beta}) + (f_{\delta} - f_{\alpha}) \exp ([v(r) - v_{\infty}]/f_{\gamma}).
\]
adopting $f_{\alpha}$ = 0.1, $f_{\beta}$ = 200 km\,s$^{-1}$, $f_{\gamma}$ = 100 km\,s$^{-1}$ and $f_{\delta}$ = 0.5 to ensure that the volume filling factor is $\sim$0.1 for the inner wind, and 0.5 in the extreme outer wind where the [O\,{\sc iii}] fine structure line originates in $\gamma$ Vel. 

To investigate the physical conditions found in the [O\,{\sc iii}] 88.4$\mu$m formation region of $\gamma$ Vel, we computed a {\sc cmfgen} model atmosphere which extends to $4.5 \times 10^6 R_\ast$, i.e. spanning $-0.4 \leq \log (N_{e}/{\rm cm}^{-3}) \leq 16.8$, requiring a significantly larger number of depth points (111) with respect to standard calculations ($\sim$50) which extend to several hundred radii. The large outer boundary necessitates adiabatic cooling, with densities closely approximating those of the ambient interstellar medium. In order to mitigate against vanishingly small populations of
high ionization stages in the outer wind, we include X-rays. We consider the impact of $f_{\delta}$ on elemental abundances in Section.~\ref{88micron}.

We have undertaken an analysis of the WC8 component of $\gamma$ Vel, updated from \citet{demarco00} which was based on the He\,{\sc i} $\lambda$5876, $\lambda$10830, He,{\sc ii} $\lambda$4686, $\lambda$5411 and C\,{\sc iii} $\lambda$6727--73, C\,{\sc iv} $\lambda$5411 diagnostics, but adjust our solution to the revised (higher) distance. The dereddened spectral energy distribution of $\gamma$ Vel from {\it IUE}, {\it ISO} SWS+LWS and {\it Herschel} PACS is compared to the theoretical WC8+O model for our adopted light ratio in Fig.~\ref{fig:wr11}. The WC8+O model reproduces observations well up to $\sim$10$\mu$m, but thereafter falls below
spectrophotometric data. 

Our spectroscopic luminosity, $\log L/L_{\odot}$ = 5.3 infers a mass of $\sim 10.6 M_{\odot}$ \citep{schaerer92} somewhat higher than the dynamically determined value of 9.0$\pm$0.6 $M_{\odot}$ from \citet{North07}, inferring $\log L/L_{\odot}$ = 5.18$\pm$0.05. 
We support the previous determination of C/He = 0.15 by number, based on He\,{\sc ii} $\lambda$5411 and C\,{\sc iv} $\lambda$5471, and consider the carbon-to-helium ratio to be relatively secure, i.e. $X_{C} = 30^{+3}_{-3}\%$. C/O = 5 by number was adopted by \citet{demarco00} since a determination of oxygen was not possible from weak/blended optical lines. Our revised physical and wind properties are derived based on the optical C\,{\sc iii-iv} diagnostics, primarily C\,{\sc iii} $\lambda$4647--51, 5696, 6727--73, 8500, plus C\,{\sc iv} $\lambda$4441, 5471, 5801--12, 7724 plus He\,{\sc ii} $\lambda$4686. Low ionization lines including C\,{\sc ii} $\lambda\lambda$6578--83, $\lambda\lambda$7231--37 and He\,{\sc i} $\lambda$5876 are somewhat underpredicted, as
is He\,{\sc i} $\lambda$1.083$\mu$m with respect to NTT-EMMI spectroscopy of \citet{demarco00}.

We present a comparison between our updated WC8 model and spectroscopic observations in Fig.~\ref{fig:spectrum}, including the combined WC8+O7.5\,III synthetic spectrum, using parameters for the O star companion updated from \citet{demarco99} to reflect the revised distance. We include a comparison with $\lambda\lambda$975-1275 {\it Copernicus} U2 spectroscopy from March 1977 in the Appendix (Fig.~\ref{fig:cop}). In the ultraviolet, the overall spectral energy distribution is well
reproduced -- iron blanketing in the O star produces the broad dip between 1500--1700\AA. Some of the prominent lines in the UV are matched satisfactorily (e.g. Si\,{\sc iv} $\lambda$1722) with respect to {\it Copernicus} and {\it IUE} spectroscopy, although others are poorly reproduced (C\,{\sc iii}] $\lambda$1909, C\,{\sc iii} $\lambda$2297) noting the complex orbital phase dependence of UV resonance lines \citep{stlouis93}. 

The predicted near-IR spectrum of $\gamma$ Vel compares qualitatively well with published near-IR spectroscopy, including C\,{\sc iii} 0.97$\mu$m, C\,{\sc iv} 2.08$\mu$m \citep{Barnes74, Aitken82}. The synthetic spectrum provides a reasonable match to C\,{\sc iv} lines within the 2.4--5$\mu$m {\it ISO}-SWS range, although strong He\,{\sc ii} lines at 4.05$\mu$m (10--8) and 4.65$\mu$m (14--10) are underpredicted. Beyond 5$\mu$m, the overall emission line spectrum is rather too weak with respect to {\it ISO}-SWS observations, especially for He\,{\sc ii} 7.45$\mu$m (6--5), C\,{\sc iii} 12.4$\mu$m (14--13), and He\,{\sc ii} 19.4$\mu$m (16-14). Overall, the C\,{\sc iii-iv} spectrum of $\gamma$ Vel is reproduced significantly better than \citet{demarco00}, albeit with He\,{\sc i} less satisfactory.

\begin{figure}
\centering
\includegraphics[width=0.4\textwidth,bb=40  45 720 585]{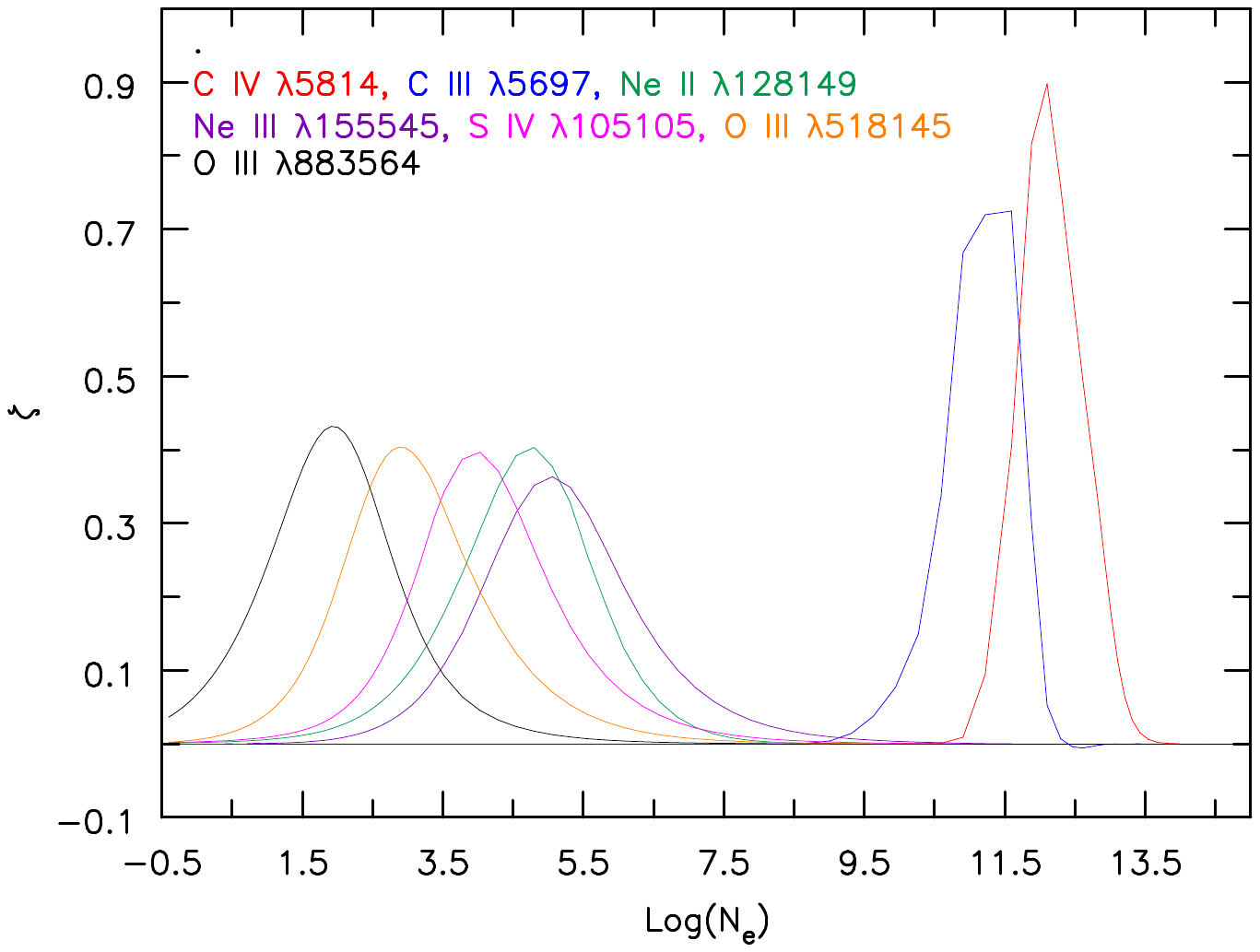}
\caption{Line formation regions of selected spectral lines in the WC8 stellar wind of $\gamma$ Vel: C\,{\sc iv} $\lambda$5812 (red), C\,{\sc iii} $\lambda$5696 (blue) in the dense inner wind, [Ne\,{\sc ii}] 12.8$\mu$m (green), [Ne\,{\sc iii}] 15.5$\mu$m (purple), [S\,{\sc iv}] 10.5$\mu$m (pink) in the low density outer wind, plus [O\,{\sc iii}] 51.8$\mu$m (orange) and 88.4$\mu$m (black) in the very low density extreme outer wind.}
\label{fig:line_form}
\end{figure}




We show the resulting wind properties as a function of electron density in 
Figure~\ref{fig:wind_struc}. Optical emission lines form within the dense inner wind ($N_{e} \sim 10^{11-12}~\mathrm{cm}^{-3}$), where the temperature is several $10^{4}$ K \citep{hillier89, hillier99}. Strong metal line cooling leads to a relatively low wind temperature of $T_{\rm e}$ = 7\,000 K for intermediate densities ($N_{e} \sim 10^{6-8}~\mathrm{cm}^{-3}$), corresponding to $\sim10^{2.5} \leq r/R_\ast \leq 10^{3.5}$. In the extreme outer wind beyond $r \sim 10^{5} R_{\ast}$  $N_{e} < 750~\mathrm{cm}^{-3}$, and  $T_{e} < 800$ K. Fig.~\ref{fig:wind_struc} also shows the  wind velocity and volume filling factor of this model, the latter corresponding to $f$=0.1 for the optical lines and $f$=0.5 for the IR fine structure lines. 


The density dependence of ionisation for carbon, oxygen, neon and sulphur, as predicted by this {\sc cmfgen} model for the WC8 star, is shown in Figure~\ref{fig:wind_struc}. 
C$^{2+}$, O$^{2+}$, S$^{+3+}$ and Ne$^{2+}$ are predicted to be the dominant ionisation stages of carbon, oxygen, sulphur and neon in the outer wind at $N_{e} \sim 10^{5}~\mathrm{cm}^{-3}$, corresponding to the fine-structure line forming regions of neon and sulphur, with O$^{2+}$ predicted to remain the dominant ion of oxygen in the extreme outer wind, where [O\,{\sc iii}] fine structure lines originate, as illustrated in Fig.~\ref{fig:line_form}. Additional elements are presented in the Appendix (Fig.~\ref{fig:wind_struc2}). He$^{+}$ remains the dominant ion of helium for all densities below $N_{e} \sim 10^{11.5}~\mathrm{cm}^{-3}$, while the dominant ions of argon and calcium in the outer wind are Ar$^{3+}$ and Ca$^{3+}$, with iron transitioning from Fe$^{4+}$ to Fe$^{3+}$ at $N_{e} \sim 10^{4.5}~\mathrm{cm}^{-3}$. 

\begin{figure*}
\centering
\includegraphics[width=0.29\textwidth,angle=-90,bb=36 200 502 570]{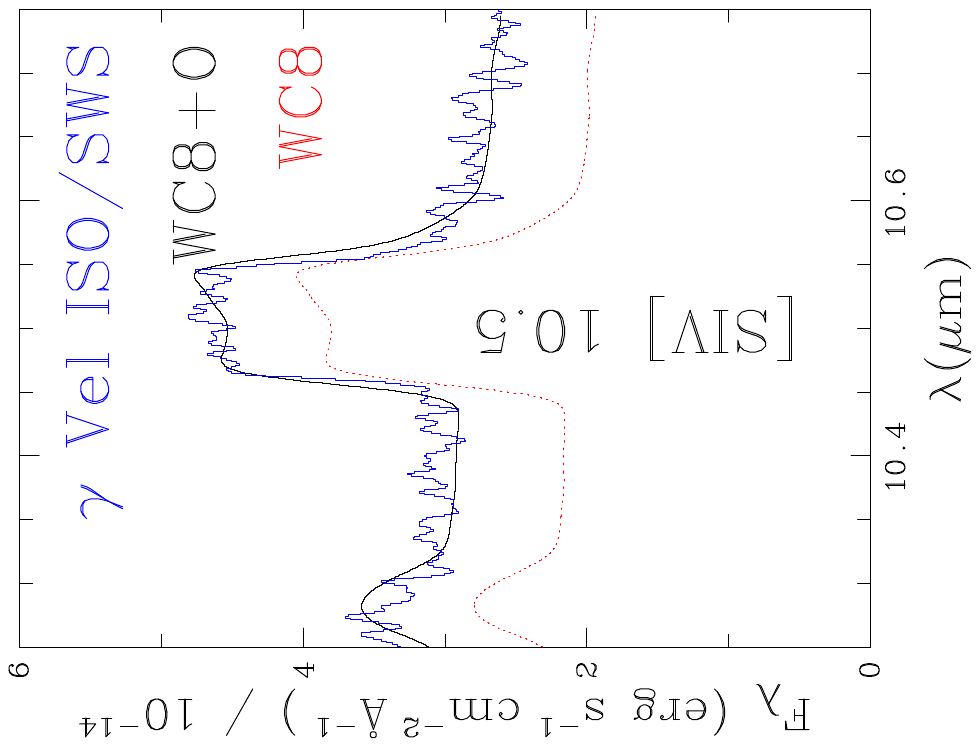}
\includegraphics[width=0.29\textwidth,angle=-90,bb=36 200 502 570]{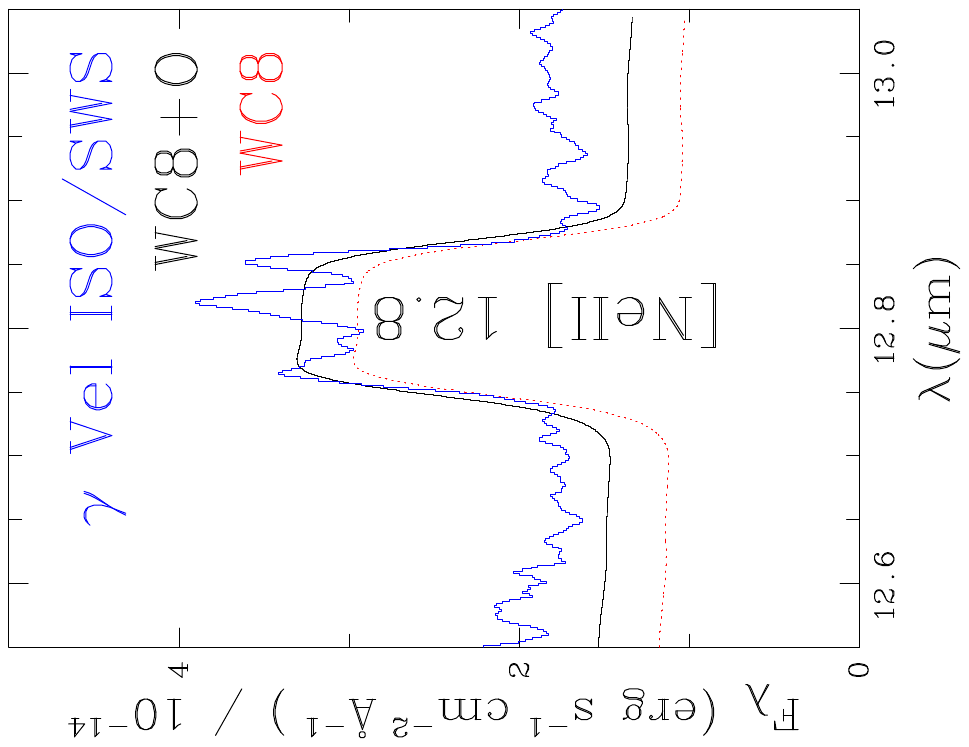}
\includegraphics[width=0.29\textwidth,angle=-90,bb=36 200 502 570]{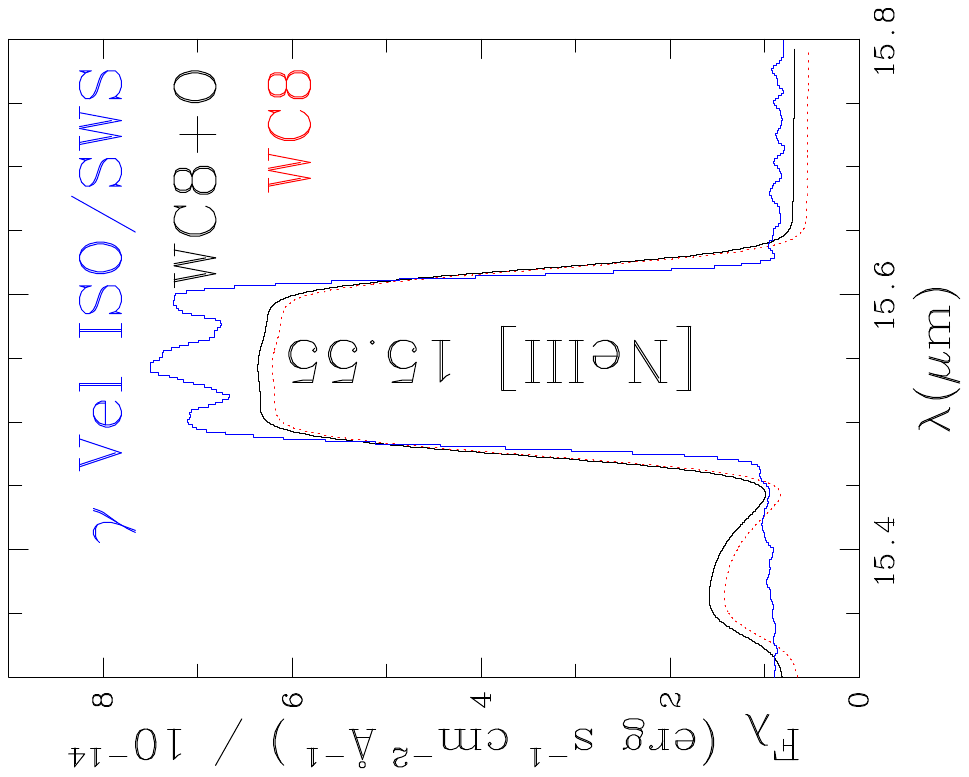}
\includegraphics[width=0.29\textwidth,angle=-90,bb=36 200 502 570]{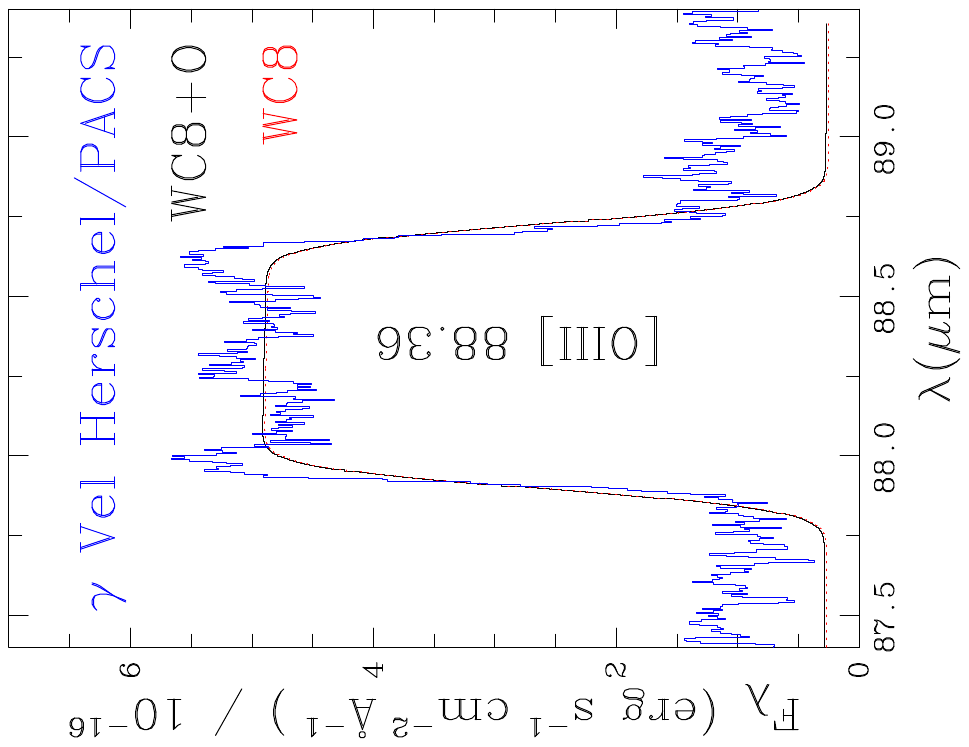}
\caption{$\gamma$ Vel's primary mid-IR fine-structure lines [S\,{\sc iv}] 10.51$\mu$m ({\it ISO}/SWS), [Ne\,{\sc ii}] 12.81$\mu$m ({\it ISO}/SWS), [Ne\,{\sc iii}] 15.55$\mu$m ({\it ISO}/SWS), [O\,{\sc iii}] 88.36$\mu$m ({\it Herschel}/PACS), shown in blue, together with the theoretical WCI+O model (black) and WC8 component (red), adjusted for instrumental broadening ($R$ = 1700, 1200, 1500, 2500, respectively).}
\label{fig:spectrum_fs}
\end{figure*}

Fig.~\ref{fig:spectrum_fs} reveals satisfactory fits to [S\,{\sc iv}] 10.5$\mu$m and [O\,{\sc iii}] 88.4$\mu$m  for mass fractions of 0.04\%, 1.0\%, respectively, plus both [Ne\,{\sc ii}] 12.8$\mu$m and
[Ne\,{\sc iii}] 15.5$\mu$m  for an adopted mass fraction of 1.7\%. Consequently,  we can be confident with the inferred oxygen, neon and sulphur abundances (for our adopted wind clumping).
[Ne\,{\sc ii}], [Ne\,{\sc iii}] and [O\,{\sc iii}] fine structure  lines show central subpeaks together with additional peaks at $\pm$1000 km\,s$^{-1}$. This structure contrasts with flat topped predictions from {\sc cmfgen}
in Fig.~\ref{fig:spectrum_fs}. 

The origin for this structure may be linked with the binary nature of $\gamma$ Vel, specifically the colliding wind region between the WC8 and O star components. The wind momentum of the WC8 component exceeds the O star by a factor of $\sim$30, so the O star wind will be strongly confined. Nevertheless the post-shock region of the colliding winds may contribute to the observed emission. Recent
 {\it James Webb Space Telescope (JWST)} MIRI imaging of the long period WC7+O binary
system WR~140 reveal nested rings of dust emission \citep{2022NatAs...6.1308L}, suggesting that a relic of its colliding winds persist to large radii.
Curiously, subpeak structure is not apparent for [S\,{\sc iv}] in $\gamma$ Vel, but would be expected from its overlapping line formation region to [Ne\,{\sc ii-iii}] (recall Fig.~\ref{fig:line_form}). Subpeaks are observed in [O\,{\sc iii}] for other binary and {\it single} WC stars (Crowther, priv. comm.), so the origin of sub-peak structures  remains uncertain.

\begin{figure}
\centering
\includegraphics[width=0.45\textwidth,bb=30 30 535 784]{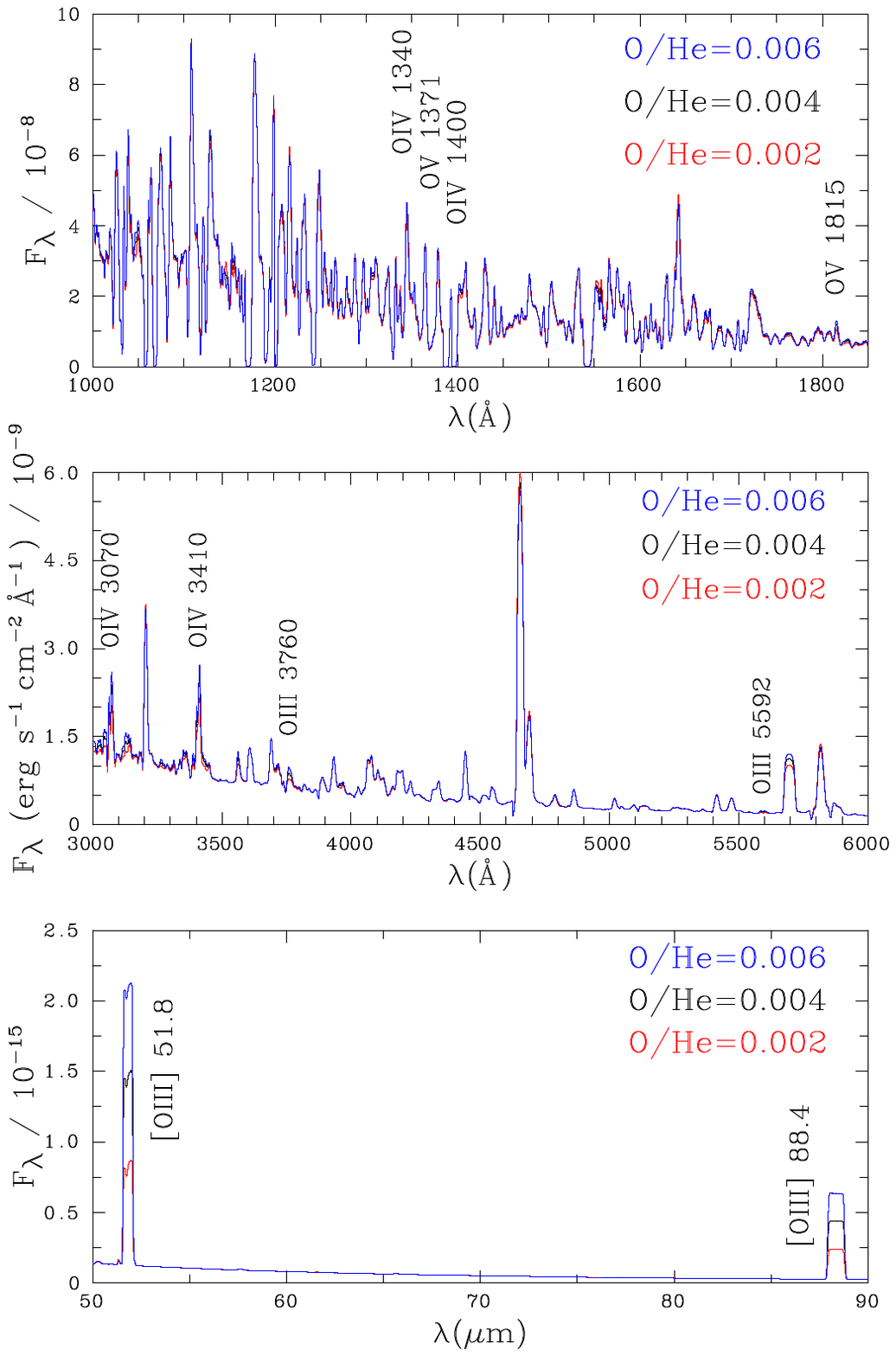}
\caption{Synthetic UV (top), visual (centre), far-IR (bottom) WC8 spectra for O/He=0.002 ($X_{\rm O}$ = 0.5\%, red), 0.004 ($X_{\rm O}$ = 1.1\%, black) and 0.006 ($X_{\rm O}$ = 1.6\%, blue) highlighting the sensitivity of fine structure lines of [O\,{\sc iii}] to oxygen abundance. UV/optical lines are relatively insensitive to a factor of three difference in oxygen content owing to the dominant helium-carbon line spectrum.}
\label{fig:oxy}
\end{figure}

Fig.~\ref{fig:oxy} compares three synthetic WC8 models in the ultraviolet, visual and far-IR with O/He = 0.002, 0.004, 0.006 by number. These illustrate the clear correlation between oxygen abundance and [O\,{\sc iii}] 52$\mu$m, 88$\mu$m strength in contrast to the few oxygen lines in the UV/optical. We will revisit our inferred abundances in Sect.~\ref{BRA88}, but favour an oxygen abundance of $\sim$1\% by mass for $\gamma$ Vel from our {\sc cmfgen} analysis, which is compared to evolutionary predictions in Sect.~\ref{summary}.

\section{$\gamma$ Vel abundances from fine structure lines}\label{BRA88}

\subsection{Oxygen}\label{88micron}

We are able to independently calculate the fractional ionic 
abundance of O$^{2+}$ from the [O\,{\sc iii}]$\,88.36\mu$m fine structure line following the approach of \citet{barlow88}, modified by \citet{dessart00} to incorporate wind clumping. 
This method was originally devised to measure ionic neon abundances in $\gamma$ Vel, and involves the [O\,{\sc iii}] line flux (from Section~\ref{obs}) and adopted wind properties (informed from Section~\ref{cmfgen}).

\begin{table}
\centering
\caption{Atomic data for [O\,{\sc iii}]$\,88.36\mu$m and $\,51.81\mu$m fine-structure lines, drawn from \citet{nussbaumer81} and \citet{Storey14} for radiative and collisional rates, respectively, with critical
densities $N_{\rm crit}$ from \citet{Rubin89}.}
\begin{tabular}{c @{\hspace{4mm}} c @{\hspace{2mm}} c @{\hspace{2mm}} c @{\hspace{2mm}} c @{\hspace{2mm}} c @{\hspace{2mm}} c @{\hspace{2mm}} c}
\hline
Transition & $\lambda$ & $\omega_u$ & $\omega_l$ & $\mathrm{A}_{ul}$ &  $\Omega_{ul}$ & $\mathrm{log}(N_{\rm crit}/\mathrm{cm}^{-3})$ \\
  $u-l$     & $\mu$m  &     &    & (s$^{-1}$) &  1\,000K &  8\,000 K   \\
\hline
$^3\mathrm{P}_2-^3\mathrm{P}_1$  &  51.814   & 5 &  3  & $9.76\times10^{-5}$  & 1.12 & 3.5 \\
$^3\mathrm{P}_1-^3\mathrm{P}_0$  &  88.356   & 3 &  1  & $2.62\times10^{-5}$  & 0.52 & 2.7 \\
\hline
\end{tabular}\par
\label{tab:atomic}
\end{table}

The following abundance determination is applicable to an emission line formed
by electronic transition between two fine-structure energy levels ($u$ and $l$),
occurring in the asymptotic region of a clumpy stellar wind, where material is on
average flowing at a velocity $v_\infty$ following a $r^{-2}$ density
distribution. If a flux $I_{ul}$ is observed in the fine structure line of a
star residing at a distance $d$, the total power emitted by the star in this
line is 
\begin{equation}
4\pi d^2I_{ul}=\int_0^\infty n_uA_{ul}h\nu_{ul}4\pi r^2 f_{\rm clump}\,\mathrm{d}r \quad {\rm erg\, s}^{-1},
\label{eq:line_power}
\end{equation}   
where $h\nu_{ul}$ is the energy of a transition with probability $A_{ul}$
($s^{-1}$) of occurrence, $n_u$ is the density of ions in the upper level, and
$f_{\rm clump}$ is the volume filling factor in the  [O\,{\sc iii}] line-forming
region. The
incorporation of this clumping factor is a modification to the original method
of \citet{barlow88}.

The density of ions in the upper level can alternatively be expressed as 
\begin{equation}
n_u=f_u n_i \quad {\rm cm}^{-3},
\label{eq:upper_density}
\end{equation}   
where $n_i$ is the species ion density, of which a fraction $f_u$ is in the upper
level. Following \citet{dessart00} we use values of $f_u$ calculated by solving
the equations of statistical equilibrium for the O$^{2+}$ ion, using the
{\sc equib} code \citep{Howarth16}, at $13$ electron densities
over the range $N_{e} = 10^0$ to $10^{12}\mathrm{cm}^{-3}$, and electron temperatures 
$T_{e}$\,{=}\,1\,000 to 10\,000 K, for the collisional and radiative atomic data presented in Table~\ref{tab:atomic}.

Knowledge of the mass-loss rate and terminal wind speed become important when
determining $n_i$. These properties are combined in the mass-loss parameter $A$
\citep[][equation 8]{barlow88} so that
\begin{equation}
n_u=\frac{f_u\gamma_iA}{f_{\rm clump} r^2} \quad {\rm cm}^{-3},
\label{eq:upper_densityA}
\end{equation}  
where $\gamma_i$ is the fraction of \emph{all} ions that are species $i$ (i.e., 
$\gamma(\mathrm{O}^{2+})$, the desired result). Combining equations \ref{eq:line_power}
and \ref{eq:upper_densityA}, the filling factor cancels and we are left with 
\begin{equation}
I_{ul}=\frac{\gamma_i}{d^2}A_{ul}h\nu_{ul}A\int_0^\infty f_u(r,f_{\rm clump},T_{e})\,\mathrm{d}r \quad {\rm erg\, s}^{-1}{\rm cm}^{-2}.
\label{eq:line_flux}
\end{equation}
At this point we follow \citet{dessart00} by integrating over electron density, $N_e$, 
removing the dependency of $f_u$ on A (and hence $\dot{M}$) and filling factor.

\setlength{\tabcolsep}{4pt}
\begin{table}
\centering
\caption{Fine-structure line fluxes  and inferred ionic abundances for $\gamma$ Vel using the integral method and a uniform filling factor of $f$ = 0.5$\pm0.25$.  
Atomic data for the transitions are given in
         Table~\ref{tab:atomic} or \citet{dessart00}, aside from updated collision strengths for
         [S\,{\sc iv}] \citep{Tayal00} and
         [Ne\,{\sc iii}] \citep{McLaughlin00}. 
         } 
\begin{tabular}{l @{\hspace{1mm}} c @{\hspace{1mm}} r @{\hspace{1mm}} c @{\hspace{1mm}} c @{\hspace{1mm}} c @{\hspace{1mm}} c @{\hspace{1mm}} c}
\hline
$\lambda$ & ID & Flux $10^{-12}$                                                  & Instrument & Ref  & $T_{e}$ & $\gamma_i$ & $X_{i}$/He \\
$\mu$m &         & erg\,s$^{-1}$\,cm$^{-2}$ & & & K & 10$^{-4}$ & 10$^{-4}$ \\
\hline
10.5        & [S\,{\sc iv}]       & 15$\pm$1        & {\it ISO} SWS & a  & 2\,000 & $\phantom{0}0.6^{+0.2}_{-0.2}$  & $0.7^{+0.2}_{-0.2}$ \\ [2pt] 
12.8        & [Ne\,{\sc ii}]      & 18$\pm$1       & {\it ISO} SWS & a &  5\,000 & $10^{+2}_{-3}$ & $12_{-4}^{+2}$ \\ [2pt] 
15.5        & [Ne\,{\sc iii}]     & 82$\pm$1        & {\it ISO} SWS & a & 5\,000 & $41_{-12}^{+9}$ & $47_{-14}^{+11}$ \\ [2pt]
51.8        & [O\,{\sc iii}]        & 8$\pm$1         & {\it ISO} LWS & b & 1\,000 & $41_{-12}^{+9}$ & $47_{-14}^{+11}$ \\ [2pt]
88.4         & [O\,{\sc iii}]       &  3.4$\pm$0.5         & {\it Herschel} PACS & b & 1\,000 & $33_{-10}^{+7}$ & $38_{-11}^{+9}$ \\ [2pt]
\hline 
\end{tabular}\par
a: \citet{dessart00}; b: this work
\label{tab:S-Ne-O}
\end{table}


Converting the integral in equation \ref{eq:line_flux} to one over $N_e$ and
re-arranging for the fractional ionic abundance, one obtains
\begin{equation}
\gamma_i=\frac{(4\pi \mu m_Hv_\infty)^{1.5}}{ln(10)}\left(\frac{\sqrt{f_{\rm clump}}}{\dot{M}^{1.5}}\right)\frac{1}{F_u(N_e,T_e)}\frac{2d^2I_{ul}}{\sqrt{\gamma_e}A_{ul}h\nu_{ul}},
\label{eq:gammai}
\end{equation}
where
\begin{equation}
F_u(N_e,T_e)=\int_0^\infty\frac{f_u(N_e,T_e)}{\sqrt{N_e}}\,\mathrm{d\,log}(N_e).
\label{eq:int_ne}
\end{equation}

Equation \ref{eq:gammai} is ultimately what we use to calculate $\gamma(\mathrm{O}^{2+}$) which is presented in Table~\ref{tab:S-Ne-O}, 
with the relevant atomic data used for their calculation given in Table \ref{tab:atomic}, adopting $T_{e}$ = 1\,000 K.
$\gamma(\mathrm{O}^{2+}) = 3.3_{-1.0}^{+0.7}\times 10^{-3}$, corresponds to $\mathrm{O}^{2+}/\mathrm{He}$ = $3.8_{-1.1}^{+0.9} \times 10^{-3}$ or $X_{\rm O}$ = 1.0$_{-0.3}^{+0.2}$\% 
by mass, assuming O$^{2+}$ is the dominant ion in the line forming region of [O\,{\sc iii}]$\,88.36\mu$m (${\sim}500~\mathrm{cm}^{-3}$).

The primary parameter affecting $\gamma\,(\mathrm{O}^{2+})$ in $\gamma$ Vel 
is wind clumping, which as discussed in Section~\ref{cmfgen}, is assumed to
persist in the extreme outer winds of WC stars with a volume filling factor
$f_{out}\,{=}\,0.5\pm 0.25$. To quantify the effect of this
uncertainty, we evaluate $\gamma\,(\mathrm{O}^{2+})$ at the extremes of this 
range, with mass-loss rates fixed at the values derived using
$f_{in}\,{=}\,0.1$. 

An alternative $\mathrm{O}^{2+}$ line is available, namely [O\,{\sc iii}]$\,51.81\mu$m from {\it ISO}-LWS (Fig.~\ref{fig:isowr11}). This line has a critical density 
an order of magnitude higher than [O\,{\sc iii}] 88.4$\mu$m (recall Table~\ref{tab:atomic}) as illustrated in Fig.~\ref{fig:line_form}.
If the ionic state of oxygen were to substantially change in the extreme outer wind one would derive different values of $\gamma(\otwop)$ using \oiii\ and 
[O\,{\sc iii}]$\,51.81\micron$, even though this is not predicted from our {\sc cmfgen} model (Fig.~\ref{fig:wind_struc}).
We derive $\gamma(\mathrm{O}^{2+})$ = $4.1_{-1.2}^{+0.9} \times10^{-3}$ from [O\,{\sc iii}] 51.81$\mu$m. 
This corresponds to $\mathrm{O}^{2+}/\mathrm{He}$ = $4.7_{-1.4}^{+1.1} \times 10^{-3}$ or $X_{\rm O}$ = $1.3_{-0.4}^{+0.2}$\% by mass, and is
in reasonable agreement with that obtained from the {\it Herschel} PACS observations (Table~\ref{tab:S-Ne-O}), despite the  low quality of the {\it ISO}/LWS spectroscopy.

Fine-structure lines of different ionisation stages of the same atomic 
species provide the best probes of ionisation structure. Unfortunately the ground state configuration of $\mathrm{O}^{+}$ does not provide a 
suitable transition, whereas the extent of $\mathrm{O}^{3+}$ may be probed using 
[O\,{\sc iv}]$~25.89\mu\mathrm{m}$ ($^2P^{\circ}_{1/2}{-}^2P^{\circ}_{3/2}$). This line has a
critical density of $N_{\rm crit} \sim 10^4\mathrm{cm}^{-3}$, so would be formed interior to both [O\,{\sc iii}]
lines considered thus far. This line is {\it not} detected in {\it ISO} SWS spectroscopy of 
$\gamma$ Vel \citep{vanderHucht96}, or indeed \textit{Spitzer} IRS scans of other
late WC stars (Crowther, priv comm), suggesting that $\mathrm{O}^{3+}$ contributes negligibly
to the outer wind of $\gamma$ Vel. Overall, fine structure lines of oxygen support the predicted outer wind structure of $\gamma$ Vel 
presented  in Figure~\ref{fig:wind_struc}.

\subsection{Neon and sulphur}\label{SNe}

We now consider fine-structure lines of 
other elements, namely neon ([Ne\,{\sc ii}] 12.8$\mu$m, [Ne\,{\sc iii}] 15.5$\mu$m) and sulphur ([S\,{\sc iv}] 10.5$\mu$m) from {\it ISO}-SWS \citep{vanderHucht96, dessart00}. [S\,{\sc iii}] 18.7$\mu$m is blended with C\,{\sc iii} 18.8$\mu$m (16--15) and He\,{\sc i} 18.6$\mu$m (3d $^{3}$D - $^{3}$P$^{\circ}$) so its detection is tentative. [Ne\,{\sc iii}] 36.0$\mu$m lies in the low sensitivity band 4 of {\it ISO}'s SWS instrument \citep{deGraauw96}, so is not measurable. One would expect $I_{36.0}/I_{15.5} \sim$ 0.04 at $T_{e}$ = 5\,000 K and $N_{e} = 10^{5}$ cm$^{-3}$ from {\sc equib} \citep{Howarth16}, i.e. $I_{36.0} \sim 3.5 \times 10^{-12}$ erg\,s$^{-1}$\,cm$^{-2}$. There is no evidence for either [Ar\,{\sc ii}] 6.98$\mu$m or [Ar\,{\sc iii}] 8.99$\mu$m in {\it ISO}-SWS spectroscopy of $\gamma$ Vel, as expected from the predicted {\sc cmfgen} ionization
balance structure of argon which suggests Ar$^{3+}$ is the dominant ion in the outer wind (Fig.~\ref{fig:wind_struc2}).  \citet{Ignace01} discuss [Ca\,{\sc iv}] 3.21$\mu$m for a sample of WR stars including $\gamma$ Vel. However, it is clear that C\,{\sc iii} 3.20$\mu$m (14--11) dominates the observed feature (their fig.~1) in the case of $\gamma$ Vel.

Using the method introduced above, together with atomic data from
\citet{dessart00} (their Table 9), we use these S and Ne line flux measurements to reassess
the fractional abundances of the respective ionic species in $\gamma$ Vel. 
These are included in Table~\ref{tab:S-Ne-O} and provide an independent insight into the empirical
ionisation balance in the outer wind versus predictions (Fig.~\ref{fig:line_form}) in which [Ne\,{\sc ii}] 12.8$\mu$m and [Ne\,{\sc iii}] 15.5$\mu$m form at
$\log(N_e/{\rm cm}^{-3}){=}\mbox{5$\pm$1} (T_{e} \sim$ 5,000 K) and [S\,{\sc iv}] 10.5$\mu$m forms at $\log(N_e/{\rm cm}^{-3}){=}\mbox{4$\pm$1} (T_{e} \sim$ 2,000 K). 

For neon,  the dominant fine structure line is [Ne\,{\sc iii}] 15.5$\mu$m, from which $\mathrm{Ne}^{2+}/\mathrm{He} = 4.7_{-1.4}^{+1.1} \times 10^{-3}$ or $X_{\rm Ne^{2+}} = 1.6_{-0.5}^{+0.3}$\% by mass
 ($T_{e} \sim$ 5,000 K is adopted), consistent with expectations for the products of the CNO cycle  being converted to $^{22}$Ne.  A neon abundance 25\% larger is obtained by including Ne$^{+}$ from [Ne\,{\sc ii}] 12.8$\mu$m, corresponding to  $X_{\rm Ne} = 2.0_{-0.6}^{+0.4}$\%, in close agreement with our {\sc cmfgen} modelling result of $X_{\rm Ne}$ = 1.7\%. Recall from Fig.~\ref{fig:wind_struc} that Ne$^{+}$ is (narrowly) the secondary ion of neon within its outer wind.
 

The inferred sulphur abundance is S/He = $0.7^{+0.2}_{-0.2} \times 10^{-4}$ by number or $X_{\rm S} \sim$ 0.04$\pm$0.01\% from [S\,{\sc iv}] 10.5$\mu$m, in good agreement with the {\sc cmfgen} modelling ($T_{e} \sim$ 2\,000 K is adopted for its line forming region). Our value is in close agreement with the solar value of $\log$ (S/H) + 12 = 7.16$\pm$0.11 \citep{2022A&A...661A.140M}, which equates to $X_{\rm S} = 0.046_{-0.010}^{+0.014}$\% by mass, as expected for an unprocessed element. In principle it could be somewhat higher if there is a non-negligible contribution from S$^{2+}$. Figure~\ref{fig:wind_struc} includes the predicted ionization structure of the WC8 outer wind of sulphur, revealing that the dominant ionization stage of sulphur is predicted to be S$^{3+}$, so one would expect [S\,{\sc iv}] 10.5$\mu$m to be dominant over  [S\,{\sc iii}] 18.7$\mu$m, as is observed.

\setlength{\tabcolsep}{4pt}
\begin{table}
\centering
\caption{Comparison between physical and wind properties of the WC8 component of $\gamma$ Vel determined here adopting the (far) interferometric distance \citep{North07} and those from \citet{demarco00} using the original (near) {\it Hipparcos} distance of \citet{1997ApJ...484L.153S}. A volume filling factor of 0.1 for the inner wind volume filling factor is adopted in both cases. Our preferred oxygen abundance is based on {\it Herschel} PACS spectroscopy of [O\,{\sc iii}] 88$\mu$m. }
\begin{tabular}{l @{\hspace{2mm}}l @{\hspace{2mm}}l @{\hspace{4mm}} l @{\hspace{2mm}} l @{\hspace{2mm}} l @{\hspace{2mm}} l @{\hspace{2mm}}l}
\hline
Study & \multicolumn{2}{c}{De Marco et al.} & \multicolumn{2}{c}{This work} \\
\hline
Distance (pc) & \multicolumn{2}{c}{258$^{+41}_{-31}$}      &     \multicolumn{2}{c}{336$^{+8}_{-7}$} \\
\hline
$T_{\ast}$/kK & 57 &   & 90 \\ [1pt]
$\log L/L_{\odot}$ & 5.0 & & 5.31 \\ [1pt]
$R_{\ast}/R_{\odot}$ & 3.2 & & 1.9 \\ [1pt]
$\log \dot{M}/M_{\odot}{\rm yr}^{-1}$ & --5.0 & & --4.84 \\ [1pt]
$v_{\infty}$ (km\,s$^{-1}$) & 1550 & & 1500$\pm$20 \\ [1pt]
$Q_{0}$ (ph\,s$^{-1}$)       & 48.81 & & 49.09 \\ [1pt]
$Q_{1}$ (ph\,s$^{-1}$)        & 47.76 & & 48.12 \\ [1pt]
C/He ($X_{\rm C}$) & 0.15 & (28\%) & 0.15$\pm$0.02 & (30$\pm$3\%) \\ [1pt]
O/He ($X_{\rm O}$) & 0.03$^{a}$ & (8\%$^{a}$) &  $0.0038_{-0.0011}^{+0.0009}$ & (1.0$_{-0.3}^{+0.2}$\%) \\ [1pt]
Ne/He ($X_{\rm Ne}$) & $^{\ddag}$0.003 &  ($^{\ddag}$1\%) &  $0.0059_{-0.0017}^{+0.0013}$ & ($2.0_{-0.6}^{+0.4}$\%) \\ [1pt]
S/He ($X_{\rm S}$) & $^{\ddag}6 \times 10^{-5}$ & ($^{\ddag}$0.03\%) & $7\pm 2 \times 10^{-5}$ & (0.04$\pm$0.01\%) \\
\hline
\end{tabular}\par
\label{summary}
$a$:O/C=0.2 by number adopted; $\ddag$: \citet{dessart00}
\end{table}

\begin{figure}
\includegraphics[width=0.45\textwidth,bb=30 350 535 784]{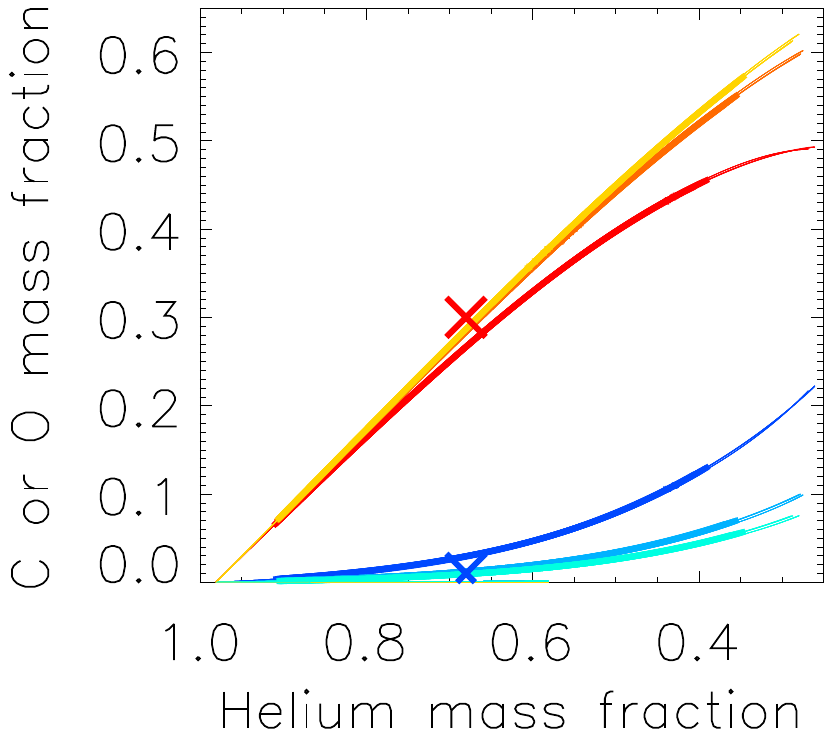}
\caption{Observed abundances by mass of carbon (red cross) and oxygen (blue cross) versus helium for $\gamma$ Vel plus predictions from BPASS v2.4 \citep{Eldridge17} for 
$M_{\rm 1,init}$ = 35--40 $M_{\odot}$, $q$= 0.5--0.9, $\log P_{\rm init}/{\rm days}$ =1--2 with $^{12}\mathrm{C}({\alpha}, {\gamma})^{16}\mathrm{O}$ reaction rates from 
\citet[][red+blue]{Angulo99}, \citet[][red+pale blue]{Kunz02} and \citet[][orange+cyan]{Kunz02} scaled by 0.75.}
\label{fig:c-vs-o}
\end{figure}

\section{Discussion and conclusions}\label{summary}

$\gamma$ Vel is a well known binary system, with a period of 78.5 days \citep{niemela80} and eccentricity of 0.33 \citep{North07},
from which component masses of 28.5$\pm$1.1 $M_{\odot}$ (O7.5\,III) and 9.0$\pm$0.6 $M_{\odot}$ (WC8) have
been established.
Binary interaction is capable of drastically altering the evolution of a massive 
star via mass and angular momentum exchange. It is likely 
that the WC8 component of $\gamma$ Vel has undergone Case B mass-transfer with its O-star 
companion \citep{eldridge09}, indicating a deviation from single-star evolution. We compare our physical and wind properties of the WC8 component of $\gamma$ Vel with previous results of \citet{demarco00} in Table~\ref{summary}. Revised values follow from a combination of the modern interferometric distance to $\gamma$ Vel, improved treatment of line blanketing  -- \citet{demarco00} solely considered Ca and Fe amongst intermediate and heavy elements -- and the availability of fine structure observations of oxygen courtesy of {\it Herschel}.

We have directly measured the oxygen abundance within the wind of $\gamma$ Vel from far-IR fine structure lines of [O\,{\sc iii}] using two complementary approaches,
detailed spectroscopic analysis with the {\sc cmfgen} model atmosphere code ($X_{\rm O} \sim$ 1.1\%), and following the analytic method devised by \citet{barlow88} and updated by \citet{dessart00} ($X_{\rm O} \sim 1.0\pm$0.3\%), Oxygen is overwhelmingly in the form of O$^{2+}$ in the outer wind of $\gamma$ Vel. Uncertain nuclear burning rates may be constrained by these quantities, as the balance between carbon and oxygen 
in massive stars is solely determined by the competition of 
$3\alpha\,{\rightarrow}\,^{12}\mathrm{C}$ and 
$^{12}\mathrm{C}({\alpha}, {\gamma})^{16}\mathrm{O}$ processes, the latter of which remains uncertain \citep{Buchmann06}. 



\setlength{\tabcolsep}{4pt}
\begin{table}
\centering
\caption{Comparison between observed (bold) and predicted BPASS \citep{Eldridge17}  v2.4  properties for $\gamma$ Vel binary for $M_{\rm 1,init}$ = 35 $M_{\odot}$, $q$= 0.8, $\log (P_{\rm init}/{\rm days}$) =1.5  \citep{eldridge09} once $X_{\rm C}/X_{\rm He}$ = 0.45 at the surface. }
\begin{tabular}{l @{\hspace{2mm}}l @{\hspace{2mm}}l @{\hspace{2mm}} l @{\hspace{2mm}} l @{\hspace{2mm}} l @{\hspace{2mm}} l @{\hspace{-2mm}}l}
\hline
$\tau$ & $M_{\rm 1}$ &  $\log L_{\rm 1}$     &  $X_{\rm He}$  & $X_{\rm C}$    & $X_{\rm O}$ & $X_{\rm Ne}$ & $^{12}\mathrm{C}({\alpha}, {\gamma})^{16}\mathrm{O}$  Rates    \\
Myr   &  $M_{\odot}$  &  $L_{\odot}$   &  \%  & \% & \% & \% &  \\
\hline
5.58   &  9.1              & 5.21                  & 64.9 & 29.2    & 3.5 & 1.9  & \citet{Angulo99}\\ 
5.57   & 9.2              &  5.23                  & 66.4 & 29.9     & 1.3 & 1.9 & \citet{Kunz02} \\ 
5.57 & 9.2                & 5.23                   & 66.6 & 30.0     & 1.0  & 1.9 &  \citet{Kunz02} $\times$ 0.75 \\ 
 \hline
$\cdots$ &  $\cdots$   &  {\bf 5.31}                  & {\bf 67}$_{-2}^{+2}$    & {\bf 30}$_{-3}^{+3}$     & {\bf 1.0}$_{-0.3}^{+0.3}$   & {\bf 2.0}$_{-0.6}^{+0.4}$   & \\ 
\hline
\end{tabular}\par
\label{tab:gvel}
\end{table}

The importance of the $^{12}\mathrm{C}({\alpha}, {\gamma})^{16}\mathrm{O}$ reaction rate in 
determining nucleosynthetic yields has been investigated by \citet{imbriani01}, 
who model the evolution a $25\,M_\odot$ star using historical cross sections for this 
reaction \citep{Caughlan85,Caughlan88}. They find that a doubling of the reaction rate in this case halves 
the carbon mass fraction at the end of helium burning, corresponding to a higher 
O/C ratio in the star as a whole. These authors conclude that the higher of their 
two considered $^{12}\mathrm{C}({\alpha}, {\gamma})^{16}\mathrm{O}$ reaction rates provides better 
agreement with the solar abundance pattern, in agreement with the earlier work of 
\citet{weaver93}. It is these studies that have motivated the choice of reaction rates 
used in many contemporary models of stellar evolution. 

However, this preference for a 
higher reaction rate is at odds with the oxygen abundances presented here. A similar issue has been
noted by \citet{Aadland22} for LMC WO stars. They provided predictions from the Geneva evolutionary model based on the NACRE reaction rates 
and NACRE rates divided by a factor of 3 (their fig.~18), which suppress the production of oxygen with respect to carbon. \citet{Aadland22}  also present
BPASS models using \citet{Kunz02} rates plus these rates scaled by 0.75, quantitatively in better agreement with our results.

In Figure \ref{fig:c-vs-o} we show C and O mass fractions versus He for BPASS v2.4 binary evolution models \citep{Eldridge17} tailored for $\gamma$ Vel. These involve a 35--40 M$_{\odot}$ primary,
$q$ = 0.5--0.9 and log $(P_{\rm init}$/days) = 1--2 at solar metallicity \citep{eldridge09}.   $^{12}\mathrm{C}({\alpha}, {\gamma})^{16}\mathrm{O}$ reaction rates are from NACRE \citep{Angulo99}, \citet{Kunz02} and
the latter scaled by 0.75 \citep{Aadland22}. Table~\ref{tab:gvel} provides a summary of predictions for different reaction rates for a 35 $M_{\odot}$ primary once $X_{\rm C}/X_{\rm He}$=0.45 (by mass) at the stellar surface. The dynamical mass
of the WC8 component from \citet{North07} is well matched for this combination of initial parameters. Our spectroscopic luminosity is a little higher, favouring a slightly higher initial (current) mass, implying a lower system age ($\sim$5 Myr). 

From Fig.~\ref{fig:c-vs-o} and Table~\ref{tab:gvel} it is apparent that either the standard or scaled \citet{Kunz02} reaction rates are favoured by our results for $\gamma$ Vel, although it is not possible to exclude modest revisions in view of uncertainties in volume filling factors. Predicted neon abundances  are also in excellent agreement with our empirical results from mid-IR fine structure lines ($X_{\rm Ne}$ = $2.0_{-0.6}^{+0.4}$\%), as anticipated for CNO nuclei processed into $^{14}$N during H burning, subsequently converted to $^{22}$Ne via the $^{14}\mathrm{N}(\alpha,\gamma)^{18}\mathrm{F}(e^{+} \nu)^{18}\mathrm{O}(\alpha,\gamma)^{22}\mathrm{Ne}$ reaction. 
The inferred sulphur abundance from mid-IR fine-structure diagnostics ($X_{\rm S}$ = 0.04$\pm$0.01\%) is in excellent agreement with the solar value \citep{2022A&A...661A.140M}.

In conclusion, we have presented successful modelling of IR fine structure lines of $\gamma$ Vel with {\sc cmfgen}, together with the complementary approach of \citet{barlow88} and \citet{dessart00}, 
suggesting oxygen and neon determinations from similar analyses of other WC (and WO) stars is realistic based on archival {\it Herschel}-PACS and {\it Spitzer}-IRAC spectroscopy, plus potentially {\it JWST}-MIRI spectroscopy of sources too faint for previous IR missions.

\section*{Acknowledgements}

Thanks to Chris Rosslowe who performed initial measurements of {\it Herschel} PACS spectroscopy of WC and WO stars. Thanks to Jan J Eldridge for providing bespoke Binary Population and Spectral Synthesis (BPASS) models (v2.4) appropriate for $\gamma$ Vel. See \citet[][v2.1]{Eldridge17}, \citet[][v2.2]{BPASS2.2} and \citet[][v2.3]{BPASS2.3} for a description of previous BPASS versions. Thanks to Alex Fullerton for converting archival {\it Copernicus} spectroscopy of $\gamma$ Vel into a user friendly format. Comments from the anonymous
referee helped to improve the manuscript. PAC and JMB are supported by the Science and Technology Facilities Council research grant ST/V000853/1 (PI. V. Dhillon). JDH acknowledges support from STScI theory grant HST-AR-16131.001-A. PR thanks the European Space Agency (ESA) and the Belgian Federal Science Policy Office (BELSPO) or their support in the framework of the PRODEX Programme.

For the purpose of open access, the author has applied a Creative Commons Attribution (CC BY) license to any Author Accepted Manuscript version arising.


\section*{Data Availability} 

The {\it Herschel} PACS spectrum will be provided on request, while the CMFGEN model of the WC8 star will be made available at https://sites.pitt.edu/~hillier/web/CMFGEN.htm


\bibliographystyle{mnras}
\bibliography{pacs_wc_paper} 

\appendix

\section{Far-UV spectroscopy of $\gamma$ Vel}

\begin{figure*}
\centering
\includegraphics[width=0.63\textwidth,angle=-90,bb=30 30 535 784]{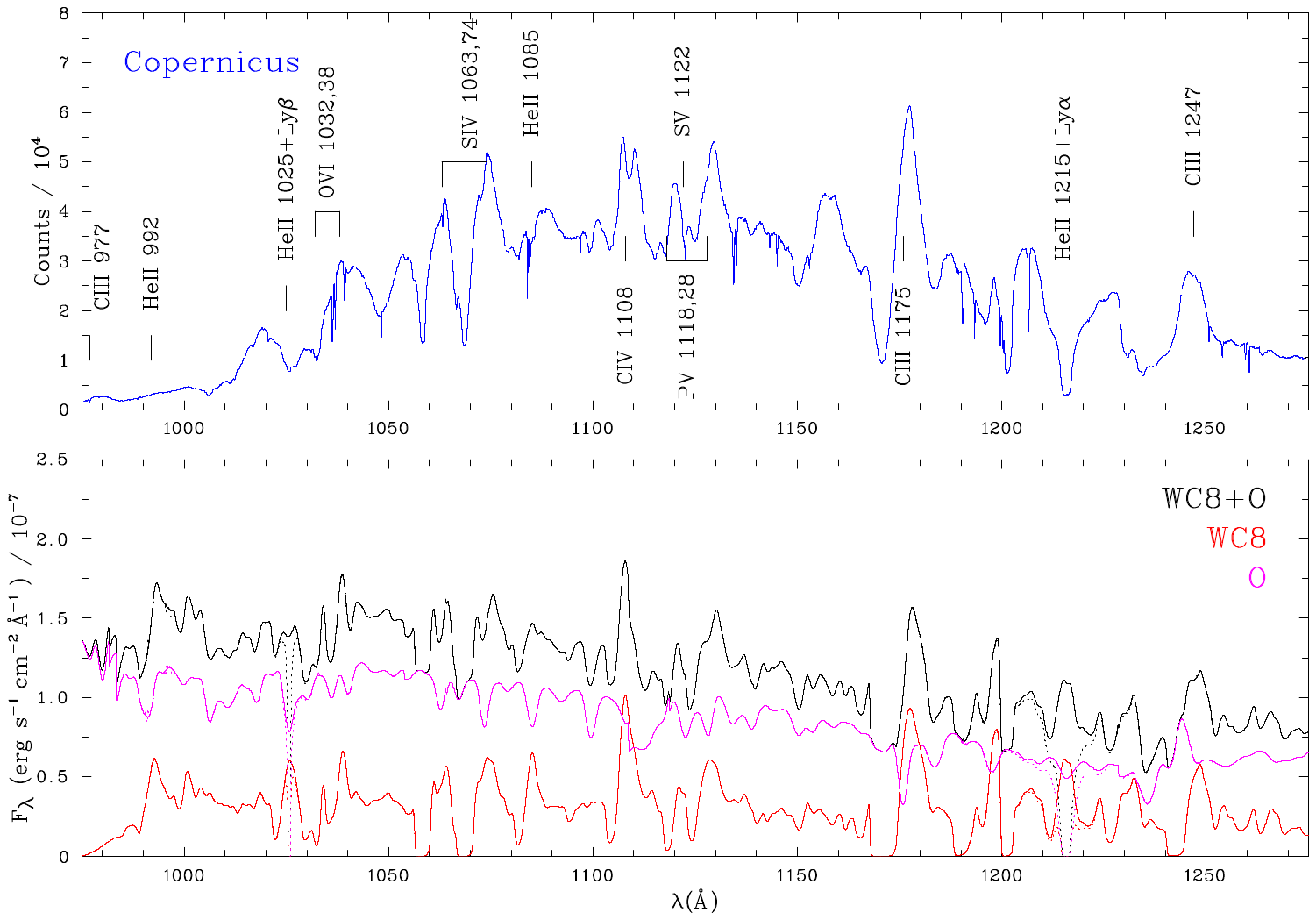}
\caption{(Top panel): Far-UV uncalibrated {\it Copernicus} U2 spectroscopy of $\gamma$ Vel covering $\lambda\lambda$975--1275 (blue), with longer wavelength datasets presented by \citet{Johnson78}; (Bottom panel): Theoretical WC8+O (black), WC8  (red) and O (pink) models, with interstellar Ly$\alpha$, $\beta$ excluded (included in dotted lines for $\log (N_{\rm HI}$/cm$^{-2}$) = 19.8 \citep{1975ApJ...200..402B, 1976ApJ...203..378Y})}
\label{fig:cop}
\end{figure*} 

\clearpage

\section{$\gamma$ Vel ionization structure}

\begin{figure*}
\centering
\includegraphics[width=0.4\textwidth,bb=40  45 720 585]{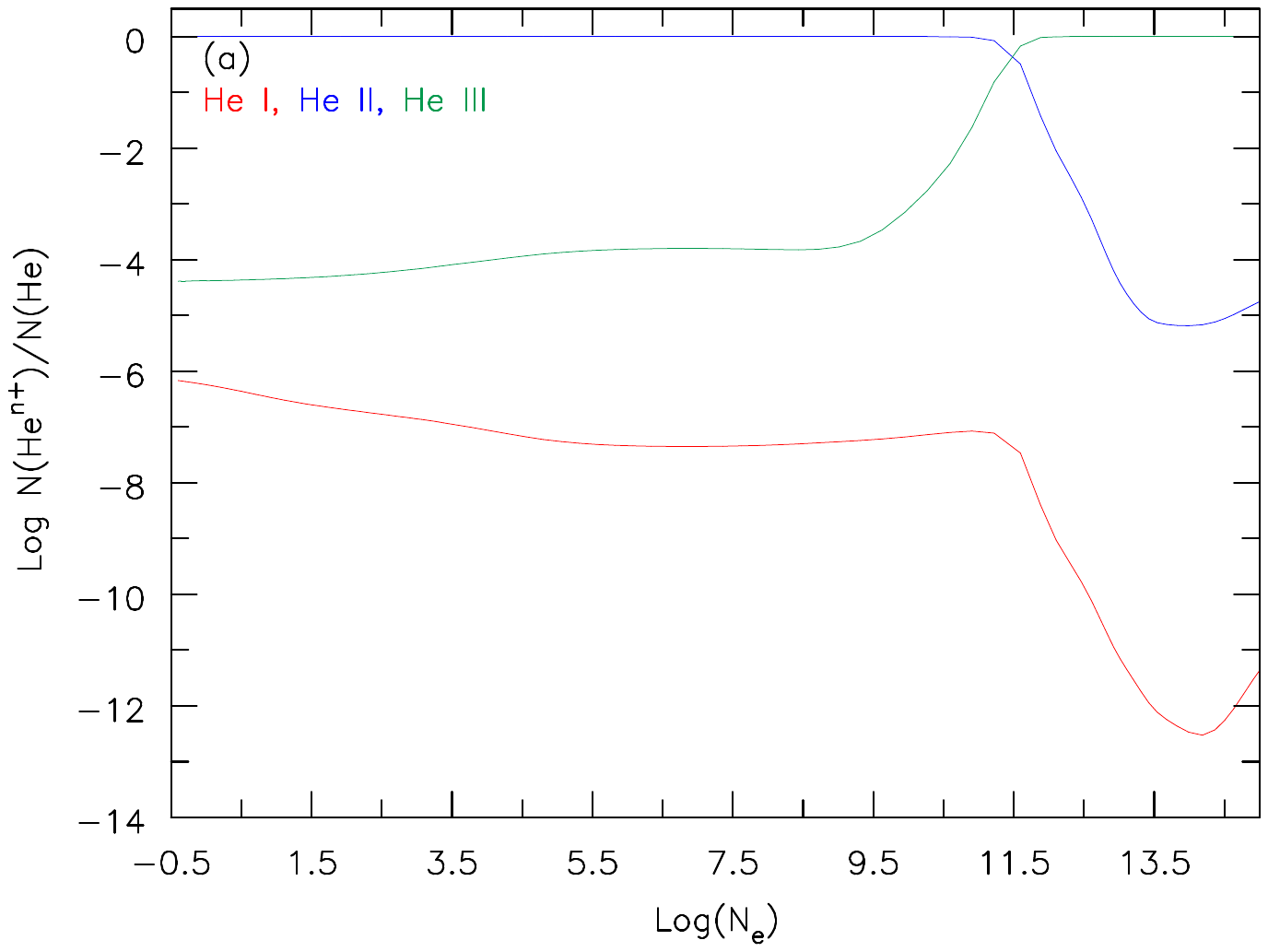}
\includegraphics[width=0.4\textwidth,bb=40  45 720 585]{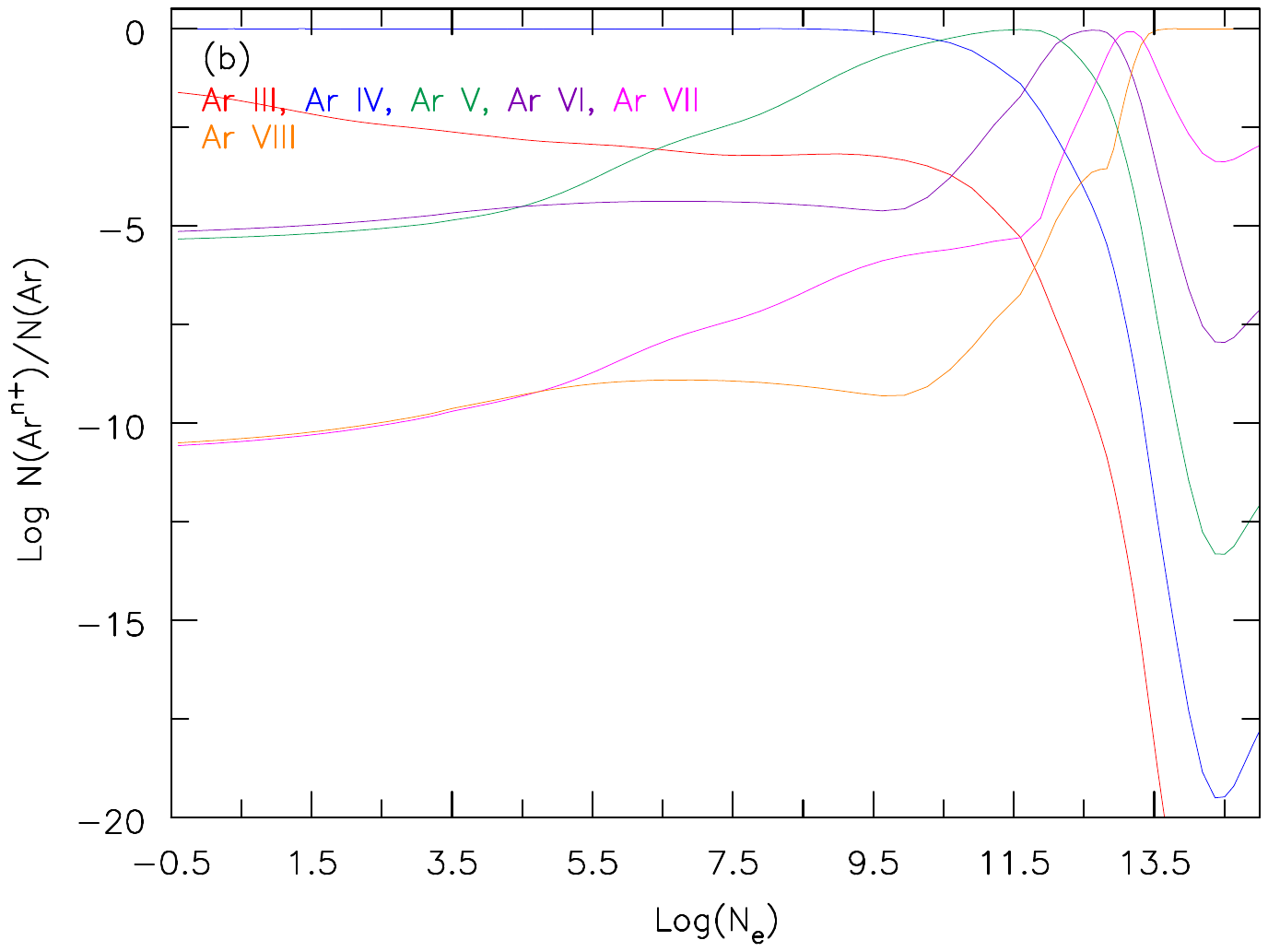}
\includegraphics[width=0.4\textwidth,bb=40  45 720 585]{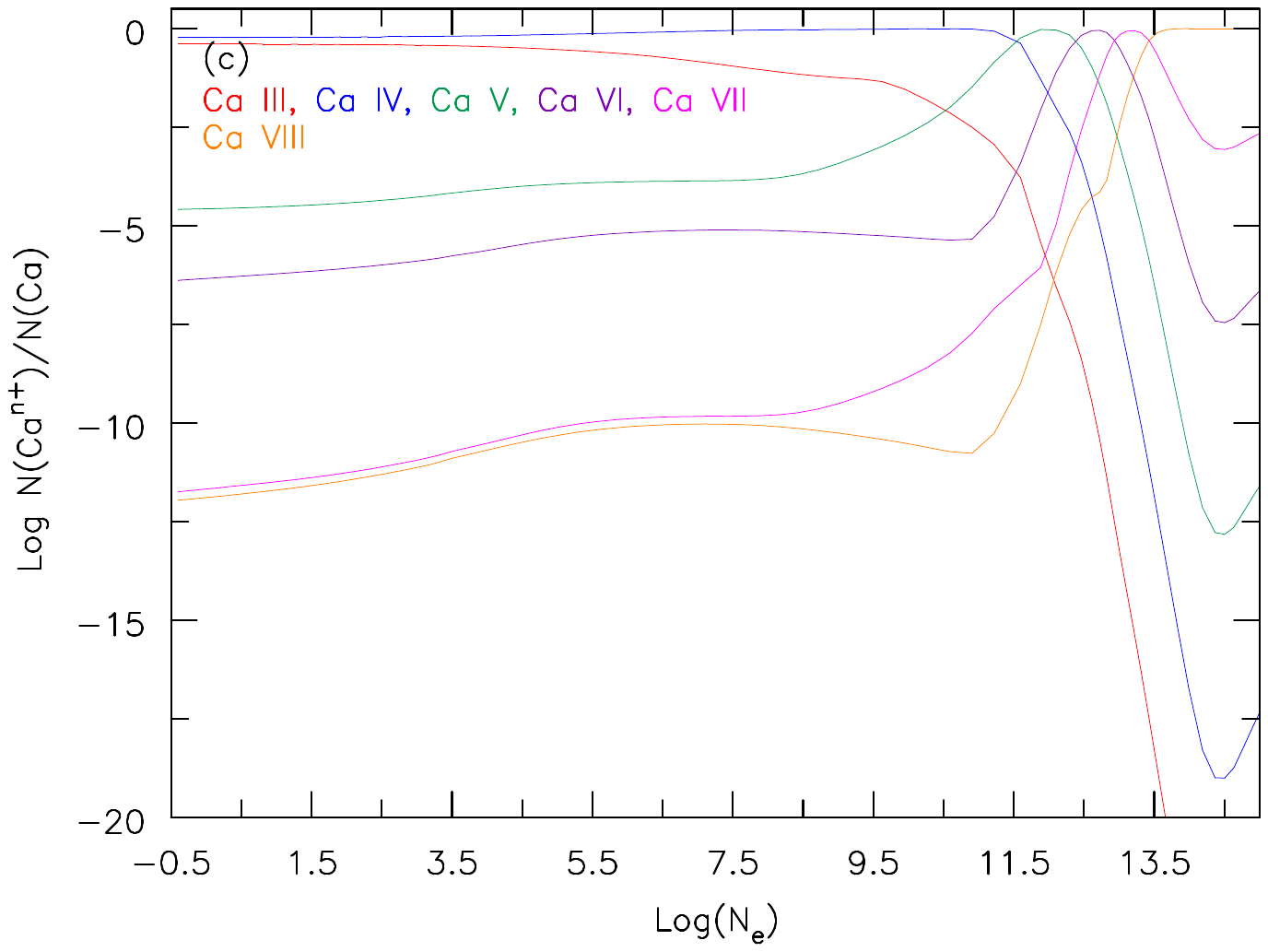}
\includegraphics[width=0.4\textwidth,bb=40  45 720 585]{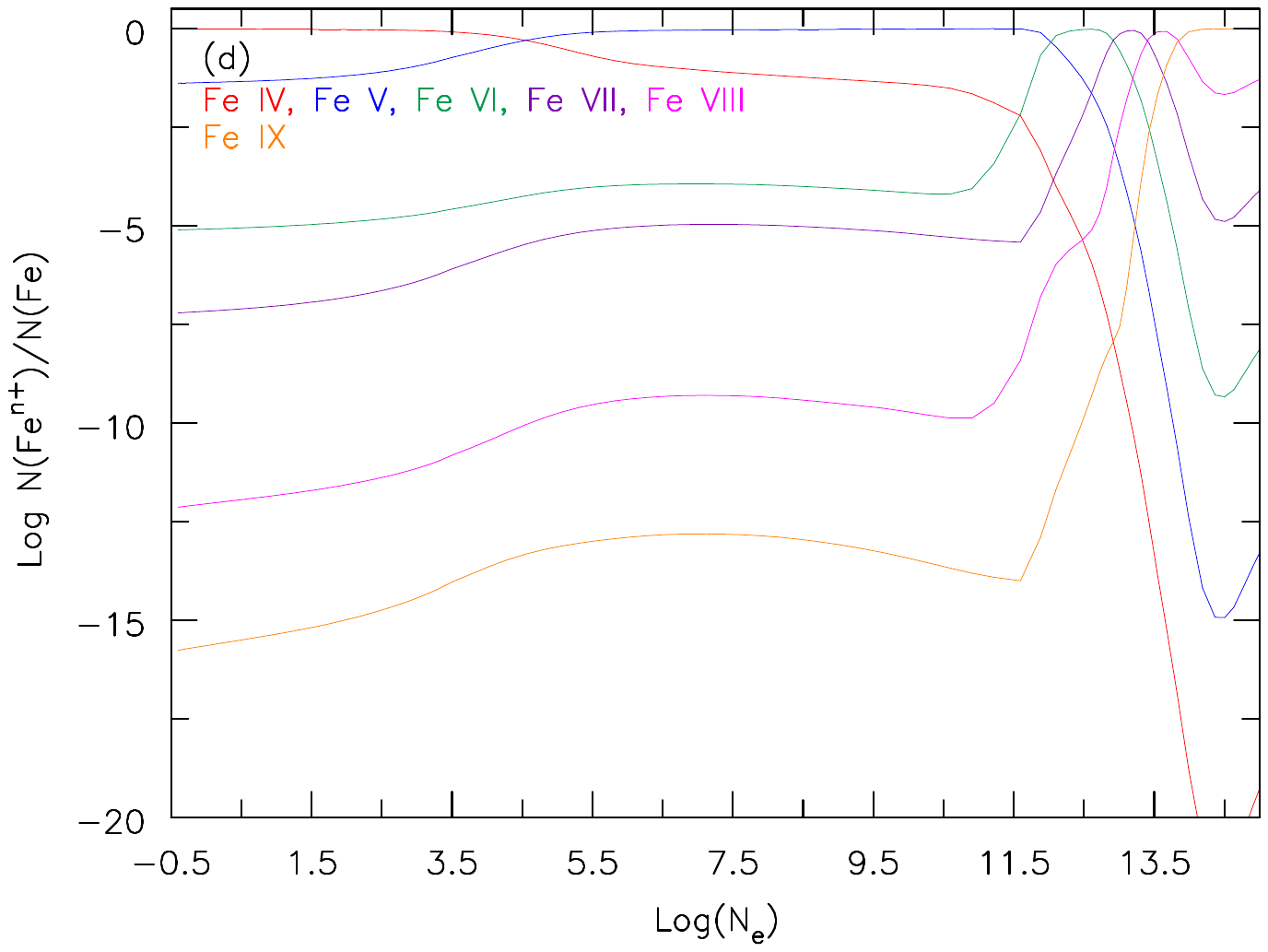}
\caption{WC8 wind structure for $\log N_{e}$ (cm$^{-3}$) versus ionization structure of (a) helium: $\log$ He$^{0}$/H (red), He$^{+}$/He (blue), He$^{2+}$/He (green); (b) argon: $\log$ Ar$^{2+}$/Ar (red), Ar$^{3+}$/Ar (blue), Ar$^{4+}$/Ar (green), Ar$^{5+}$/Ar (purple), Ar$^{6+}$/Ar (pink), Ar$^{7+}$/Ar (orange); (c) calcium: $\log$ Ca$^{2+}$/Ca (red), Ca$^{3+}$/Ca (blue), Ca$^{4+}$/Ca (green), Ca$^{5+}$/Ca (purple), Ca$^{6+}$/Ca (pink), Ca$^{7+}$/Ca (orange); (d) iron: $\log$ Fe$^{3+}$/Fe (red), Fe$^{4+}$/Fe (blue), Fe$^{5+}$/Fe (green), Fe$^{6+}$/Fe (purple), Fe$^{7+}$/Fe (pink), Fe$^{8+}$/Fe (orange). Higher ionization stages will dominate optically thick regions at high densities ($\log N_{e} \geq 13.5$).}
\label{fig:wind_struc2}
\end{figure*}








\bsp	
\label{lastpage}
\end{document}